\documentclass[aps,twocolumn,preprintnumbers,nofootinbib,superscriptaddress]{revtex4}
\usepackage{eurosym}
\usepackage{amsmath}
\usepackage{graphicx}
\usepackage{amsfonts}
\usepackage{array}
\usepackage{amsthm}
\usepackage{bm}
\usepackage{palatino}
\usepackage{mathpazo}
\usepackage{supertabular}
\usepackage[colorlinks]{hyperref}
\usepackage{subfig}
\usepackage[font=small,labelfont=bf,justification=justified,format=plain]{caption}
\usepackage[font=small,labelfont=bf,justification=justified]{caption}
\usepackage{color}
\usepackage{booktabs}
\usepackage{multirow}
\usepackage[labelfont=bf,labelsep=quad,justification=justified]{caption}
\usepackage{xcolor}
\newcommand{\be}{\begin{equation}}
\newcommand{\ee}{\end{equation}}
\newcommand{\ba}{\begin{eqnarray}}
\newcommand{\ea}{\end{eqnarray}}
\newcommand{\bal}{\begin{align}}
\newcommand{\eal}{\end{align}}

\newcommand{\bw}{\begin{widetext}}
\newcommand{\ew}{\end{widetext}}

\newcommand{\va}{\varepsilon}

\begin{document}
\title{Rotating regular black holes in conformal massive gravity}
\author{Kimet Jusufi}
\email{kimet.jusufi@unite.edu.mk}
\affiliation{Physics Department, State University of Tetovo, Ilinden Street nn, 1200,
Tetovo, North Macedonia}
\affiliation{Institute of Physics, Faculty of Natural Sciences and Mathematics, Ss. Cyril
and Methodius University, Arhimedova 3, 1000 Skopje, North Macedonia}
\author{Mubasher Jamil}
\email{mjamil@zjut.edu.cn (corresponding author)}
\affiliation{Institute for Theoretical Physics and Cosmology, Zhejiang University of Technology, Hangzhou, 310023, China}
\affiliation{Department of Mathematics, School of Natural
	Sciences (SNS), National University of Sciences and Technology
	(NUST), H-12, Islamabad, 44000 Pakistan}
\affiliation{United Center for Gravitational Wave Physics (UCGWP), Zhejiang University of Technology, Hangzhou, 310023, China}

\author{Hrishikesh Chakrabarty}
\email{chrishikesh17@fudan.edu.cn}
\affiliation{Center for Field Theory and Particle Physics and Department of Physics, Fudan University,
2005 Songhu Road, Shanghai, China}

\author{Qiang Wu}
\email{wuq@zjut.edu.cn}
 \affiliation{Institute for Theoretical Physics and Cosmology, Zhejiang University of Technology, Hangzhou, 310023, China}

\author{Cosimo Bambi}
\email{bambi@fudan.edu.cn}
\affiliation{Center for Field Theory and Particle Physics and Department of Physics, Fudan University,
	2005 Songhu Road, Shanghai, China}
	
\author{Anzhong Wang}
\email{Anzhong\_Wang@baylor.edu}
\affiliation{GCAP-CASPER, Physics Department, Baylor University, Waco, TX 76798-7316, USA}
\affiliation{Institute for Theoretical Physics and Cosmology, Zhejiang University of Technology, Hangzhou, 310023, China}

\begin{abstract}
In this paper, we use a suitable conformal rescaling to construct static and rotating regular black holes in conformal massive gravity. The new metric is characterized by the mass $M$, the  ``scalar charge'' $Q$,  the angular momentum parameter $a$, the ``hair parameter'' $\lambda$, and the conformal scale factor encoded in the parameter $L$. We explore the shadow images and the deflection angles of relativistic massive particles in the spacetime geometry of a rotating regular black hole.
For $\lambda \neq 0$ and $Q > 0$, the shadow is larger than the shadow of a Kerr black hole. In particular, if  $\lambda < 0$, the shadow radius increases considerably. For $\lambda \neq 0$ and $Q < 0$, the shadow is smaller than the shadow of a Kerr black hole. Additionally we put observational constraints on the parameter $ Q $ using the latest Event Horizon Telescope (EHT) observation of the supermassive black hole M87*.
Lastly, using the Gauss-Bonnet theorem, we show that the deflection angle of massive particles is strongly affected by the parameter $L$. The deflection angle might be used to distinguish rotating regular black holes from rotating singular black holes. 
\end{abstract}

\maketitle
\section{Introduction} 

Einstein's general theory of relativity is the current framework to describe the geometrical structure of the spacetime and the gravitational dynamics of massive bodies. This theory has been extensively tested in the weak field regime and we are now entering an era of precision experiments with gravitational wave detectors and electromagnetic observations which would make it possible to explore gravity in the highly nonlinear dynamical regime \cite{Will,r1,r2,r3}. 

One of the most striking predictions in this theory is the existence of black holes. Black holes are spacetime regions where the gravitational field is so strong that nothing, not even light, can escape. According to Einstein's theory, at the center of a black hole, there is a gravitational singularity. Despite the efforts and the great attention in this direction, the problem of black hole singularity has not been solved yet. On the other hand, it is widely believed that the gravitational field at a much deeper level should be described by a quantum theory in terms of a spin-2 particle known as gravitons.  Additionally, the scenario of a massive graviton has been considered by many authors, for example, to solve the hierarchy problem. In the Refs.~\cite{Dvali,Dvali1}, it has been argued that the brane-world gravity scenarios suggest a massive gravity. Historically, the idea of massive gravity was first investigated by Fierz and Pauli \cite{Fierz}.
It has been shown that the theory of massive gravity is not free from ambiguities, for example, we can mention here the existence of vDVZ (van Dam-Veltman-Zakharov) discontinuity. To resolve this problem, Vainshtein introduced a mechanism \cite{Vainshtein} to avoid the discontinuity problem. Yet another problem associated with this theory, was the ghost instability at the non-linear level \cite{Boulware}. Later on, to avoid such an instability, de Rham, Gabadadze and Tolley (dRGT) \cite{Rham}  proposed a new massive gravity theory by extending the Fierz-Pauli theory. In this direction, other models have been proposed \cite{Bergshoeff,Hassan}.
Recently, the black hole thermodynamics has been studied in dRGT massive gravity \cite{Cai}, while in Ref.~\cite{Katsuragawa} a few authors studied neutron stars in the context of massive gravity. More recently, Bebronne and Tinyakov \cite{bebronne} obtained spherically-symmetric vacuum solution in massive gravity. This solution is shown to depend on the mass  $M$, a quantity known as the scalar charge $Q$, and a parameter $\lambda$. It was used in \cite{Capela} to study the validity of the laws of thermodynamics. Finally, let us point out that other solutions in massive gravity have been reported in literature \cite{Xu}.

In this paper, we present the shadow cast by black holes in massive gravity. First, we consider the spherically symmetric black hole solution of \cite{bebronne} and we show that this solution is non-singular in a theory with conformal symmetry. Later, using a modified Newman-Janis algorithm we obtain a rotating solution in massive gravity which is also non-singular in the theory with conformal symmetry. The conformal invariance is expected to obtain via a theory where the massive gravity metric $ g_{\mu\nu} $ is replaced by an auxiliary dilaton field $ \Phi $ and the metric $ \hat{g}_{\mu\nu} $ is given by
\begin{equation}
g_{\mu\nu} = \left( \Phi^2\kappa^2_4 \right)\hat{g}_{\mu\nu},
\end{equation}
where $ 2/\kappa^2_4 = \frac{1}{16\pi G_N} $. However, the world around us is not conformally invariant and we must thus find a mechanism to end up with a low energy effective action without conformal symmetry. For instance, the symmetry may be spontaneously broken, or it may be realized in the UV regime at a UV fixed point \cite{Bambi2}. At high energies, when conformal invariance is restored, there are no mass scales. So the mass of the graviton should appear when the conformal symmetry is broken. There are different realizations, but in the simplest scenario we will have a dilaton field $ \Phi $ that gives both the mass of the graviton and the value of Newton's constant, and we may thus need a small constant to link the two quantities. 

One of our goals here is to study the shadows of both non-rotating and rotating regular black holes in massive gravity. Shadows possess interesting observational signatures and, in the future, it may be possible to put observational constraints on gravitational theories from shadow observations. While the characteristic map of a shadow image would depend on the details of the astrophysical environment around the black hole, the shadow contour is determined only by the spacetime metric itself. In light of this, there have been efforts to investigate shadows cast by different black holes and compact objects. The shadow of a Schwarzschild black hole was first studied by Synge \cite{Synge66} and Luminet \cite{Luminet79} and the same for Kerr black hole was studied by Bardeen \cite{DeWitt73}. Since then various authors have studied shadows in modified theories of gravity and wormholes \cite{Zakharov05,Stuchlik:2019uvf,Shipley:2016omi,Gott:2018ocn,Takahashi:2005hy,Guo:2018kis,Mureika:2016efo,Moffat:2015kva,Hioki:2008zw,Li:2013jra,Abdujabbarov:2016hnw,Amir:2016cen,Saha:2018zas,Abdujabbarov:2012bn,Ayzenberg:2018jip,Cunha:2016wzk,Atamurotov:2013dpa,Atamurotov:2013sca,Bambi:2019tjh,Vagnozzi:2019apd,Jusufi:2019nrn,Zhu:2019ura,Haroon:2018ryd,Amir:2018pcu,Shaikh:2018kfv,Shaikh:2018lcc,Gyulchev:2019osj,Gyulchev:2018fmd,Haroon:2019new,Abdujabbarov:2015pqp,Bambi:2008jg,Bambi:2010hf,Abdikamalov19,Zhou19}. On the other hand, gravitational deflection by black holes is an interesting topic. For example, one can use it to distinguish different spacetime geometries. Some recent contributions to the problem of the deflection of massive particles can be found in Refs.~\cite{Crisnejo:2019ril,Crisnejo2,Crisnejo3,Crisnejo4,Jusufi1,Jusufi2,Tsupko:2014wza,Jia,Li:2019pvi,Pang,kerr1}. Note that photon trajectories are independent of the conformal factor, so our shadow calculations are to test the black hole solutions in massive gravity, not the conformal factor in the metric.

This paper is structured as follows. In Sec. \textbf{II}, we review the black hole solution in massive gravity. In Sec. \textbf{III}, we construct a static regular black hole solution in massive gravity. In \textbf{IV}, we extend the static solution to a rotating solution. In \textbf{V}, we construct a singularity-free rotating black hole in massive gravity. In Secs.\textbf{VI} and \textbf{VII}, we study the geodesics equations and the shadow images of these black holes. In \textbf{VIII}, we consider the problem of gravitational deflection of massive particles. Finally, in Sec. \textbf{IX}, we comment on our results. 

\section{Black  holes in massive gravity}

Let us begin with a brief review of the black hole solution in massive gravity.
Our massive gravity theory is described by the action \cite{Dubovsky}
\begin{equation} \mathcal{S}_{MG} = \int  d^4x  \sqrt{-g} \left[ \frac{R}{16 \pi }+ \Lambda^4 \mathcal{F}(X, W^{ij}) \right] ,
\end{equation}
where $R$ is as usual the scalar curvature and $\mathcal{F}$ is a function of the scalar fields $\psi^i$ and $\psi^0$,  which  are minimally coupled to gravity.  These scalar fields play the crucial role  for spontaneously breaking Lorentz
symmetry.  Actually, this  action in  massive gravity  can be treated  as the  low-energy effective theory below the ultraviolet cutoff   $\Lambda$.
The value of $\Lambda$ is of  the order of $\sqrt{mM_{pl}}$ ,  where $m$ is the graviton
mass and $M_{pl}$ is  the Planck mass.
The function $\mathcal{F}$ depends on two particular combinations of the derivatives of the Goldstone fields, $X$ and $W^{ij}$, which are defined as
\begin{equation} X = \frac{\partial^0 \psi^i\partial_0\psi^i}{\Lambda^4},
\end{equation}
\begin{equation}
W^{ij} = \frac{\partial^\mu \psi^i\partial_\mu \psi^j}{\Lambda^4}-\frac{\partial^\mu \psi^i\partial_\mu \psi^0 \partial^\nu \psi^j\partial_\nu \psi^0}{\Lambda^4 X},
\end{equation}
where the constant $\Lambda$ has the dimension of mass. From this, one can arrive at
a new type of black hole solutions, namely, massive gravity black holes (detailed derivation can be found in \cite{bebronne}).
The ansatz for the static spherically symmetric black hole solutions can be written in the following form:
 \begin{equation} \label{metric1}
ds^2=- f(r) dt^2+\frac{dr^2}{g(r)}+r^2\left(d\theta^2+\sin^2\theta d\phi^2   \right),
\end{equation}
where the metric function with the scalar fields are assumed in the following form
  \begin{equation} \label{metric1}
f(r)=g(r)=  1-\frac{2M}{r}-\frac{Q}{r^{\lambda}}, 
\end{equation}
and
\begin{equation} \label{metric1}
\psi^0 = \Lambda^2(t+N(r) ),\,\,\,\, \psi^i= \Lambda^2x^i,
\end{equation}
with
 \[N(r) = \pm \int\frac{dr}{f(r)} \left[ 1-f(r) \left( \frac{Q\lambda (\lambda-1)}{12m^2} \frac{1}{r^{\lambda+2}} +1\right)^{-1}\right]^{\frac{1}{2}},\]
where $M$ is the gravitational mass of the body and
$\lambda$ is a parameter of the model that depends on the scalar charge $Q$. The presence of the
scalar charge modifies the Schwarzschild  solution in an interesting way. 

In the present article, we shall consider $M > 0$ along with the possibilities: $Q > 0$ and $Q<0$, $\lambda>0$ and $\lambda<0$, respectively. 
From the spacetime metric, one can easily observe its deviation from the usual Schwarzschild black hole due to the presence of the scalar charge $Q$ and the ``hair parameter'' $\lambda$. Consequently one can show that the attractive gravitational potential can be stronger or weaker than the usual Schwarzschild black hole depending on the sign before $Q$.

\section{Regular black holes in conformal massive gravity}

In this section, we construct a singularity-free spherically symmetric black hole in massive gravity. In \cite{Bambi1,Bambi2,Chakrabarty17}, the authors presented a method to find regular black hole solutions by rescaling the metric with a scale factor. The resultant metric becomes a solution of a theory having conformal symmetry and is regular everywhere. The underlying idea behind conformal gravity is that the spacetime singularities appearing in gravity theories are just an artifact of gauge in conformal theory and they can be removed by a suitable gauge transformation, which in this case is the rescaling that we perform \cite{Bambi1,Bambi2,Chakrabarty17,Modesto1,Rachwal1,Toshmatov1}. 

Here we also check the regularity of the spacetime by studying preliminarily the curvature invariants, and in detail the geodesic completion of massive and massless particles. However, we need to emphasize that the scalar curvatures are invariants in Einstein's gravity but they are not co-covariant in conformal gravity (invariant under both Weyl and general coordinate transformations). Therefore they cannot be used to check the regularity of spacetime and we need to rely on geodesic completion. 

In conformal gravity, the theory is invariant under both coordinate and conformal symmetries,
\begin{equation}
	\begin{aligned}
		& x^\mu \rightarrow x'^\mu (x^\mu), \\
		& g^{\mu\nu} \rightarrow g'^{\mu\nu} = \Omega^2g^{\mu\nu}.
	\end{aligned}
\end{equation}
From the previous work, we expect that the conformal factor $ \Omega $ capable of resolving the singularity should be singular at the singularity of the metric. 

\begin{figure*}[t]
	\begin{center}
		\includegraphics[type=pdf,ext=.pdf,read=.pdf,width=7.0cm]{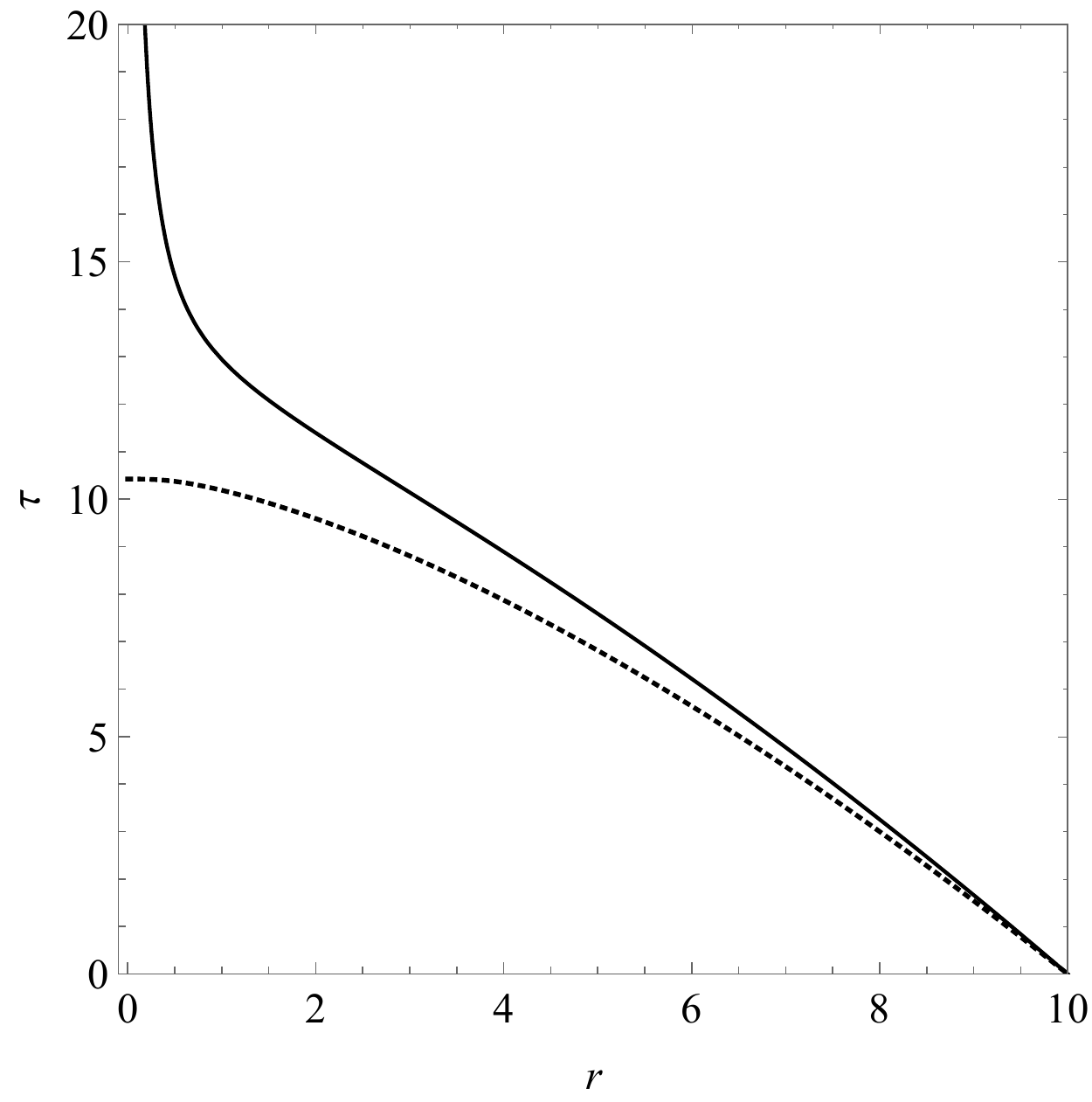}
		\includegraphics[type=pdf,ext=.pdf,read=.pdf,width=7.0cm]{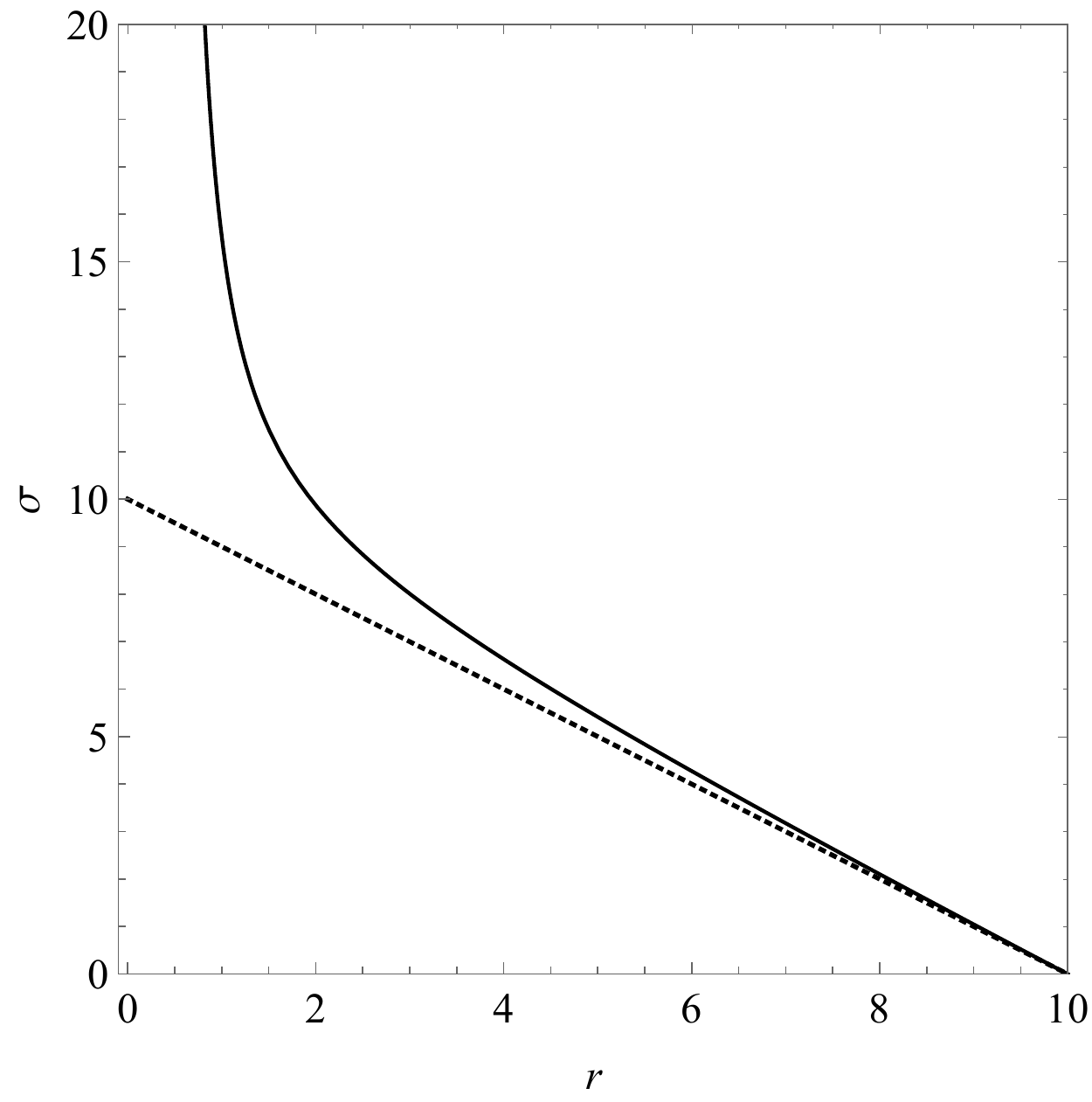}
	\end{center}
	\vspace{-0.5cm}
	\caption{Left panel: proper time $\tau$ as a function of the radial coordinate $r$ of a massive particle with vanishing angular momentum moving to smaller radii from the initial coordinate $r = r_{\rm in}$. The solid line corresponds to the proper time in the rescaled metric and the particle cannot reach the surface $r = r_{\rm sing}$. The dashed line corresponds to the standard metric in massive gravity. Right panel: as in the left panel for the affine parameter $\sigma$ of a massless particle. In these plots, we assume $E=L=1$, $ M = 2 $, $ \lambda = 4 $, and $r_{\rm in}=10$. \label{geodesic_a}}
\end{figure*}

We now explicitly provide -- in whatever conformally invariant theory -- an example of a singularity-free exact black hole solution obtained by rescaling the Schwarzschild metric by a suitable overall conformal factor $\Omega$. The new singularity-free black hole metric takes the form
\begin{equation}\label{NRBH} 
	ds^{* 2} \equiv 
	\hat{g}_{\mu\nu}^* dx^\mu dx^\nu = W(r) \hat{g}_{\mu\nu} dx^\mu dx^\nu, 
\end{equation}
where
\begin{equation}
	W(r) = \Omega^2  = \left( 1 + \frac{L^2}{r^2} \right)^{\frac{|\lambda|}{2}+2}.
\end{equation}
Here $\lambda$ is the parameter of the model of massive gravity and $ L $ is a parameter introduced for dimensional reasons.  It could be equal to the Planck length $L = L_P$, the fundamental scale of the theory, or even $L \propto M$. The scale factor $W(r)$ meets the conditions $W^{-1}(0) =0$ and $W^{-1}(\infty) = 1$. Moreover the singularity (at $r=0$) appears exactly where the conformal transformation becomes singular, i.e. where $W^{-1} = 0$. Here we must understand the singularity issue as an artifact of the conformal gauge. In \cite{Zhou19} the authors have reported a  constraint of $ L/M < 0.12 $ for a regular Kerr black hole from the analysis of a 30 ks  \textit{NuSTAR} observation of the stellar-mass black hole in GS 1354–645 during its outburst in 2015. 

In general, one can have infinite class of such functions $W(r)$ that enables us to map the singular Schwarzschild spacetime to an ``everywhere regular" one. The metric after conformal rescaling can be written as
\begin{eqnarray}
	ds^{*2} &=&  - \left(1 + \frac{L^2}{r^2}\right)^{\frac{|\lambda|}{2}+2}\left( 1 - \frac{2M}{r} -\frac{Q}{r^\lambda}\right) dt^2 \\\notag
	&+&  \left(1 + \frac{L^2}{r^2}\right)^{\frac{|\lambda|}{2}+2}\frac{dr^2}{1 - \frac{2M}{r}-\frac{Q}{r^\lambda} } \\ \notag
	&+&
	\left(1 + \frac{L^4}{r^2}\right)^{\frac{|\lambda|}{2}+2} r^2 \left( d\theta^2 + \sin^2\theta d\phi^2  \right)  \, .  
\end{eqnarray}
For the metric above, the Kretschmann invariant ${ \hat{K} } = {\bf \hat{R}iem}^2$ and the Ricci scalar $ \hat{\mathcal{R}} $ are reported in the Appendix \ref{apend_a}. These scalar invariants are regular everywhere in the spacetime including $ r = r_{sing} $. 

Now we would like to consider the regularity of spacetime we obtained by studying the geodesic motion of massive and massless particles.

For massive particles, we have $ \hat{g}^*_{\mu\nu}\dot{x}^\mu\dot{x}^\nu  = -1 $, where the dotted quantities denote the derivative with repect to the proper time $ \tau $. In this analysis, we consider purely radial geodesics only,  i.e. $ \dot{\theta} = \dot{\phi} = 0 $. Hence the equation becomes
\begin{equation}\label{massivemaster}
	\hat{g}^*_{tt}\dot{t}^2 + \hat{g}^*_{rr}\dot{r}^2 = -1.
\end{equation} 
The metric under consideration is independent of time coordinate, therefore we  have the conservation of the particle energy $ E $
\begin{equation}\label{energycon}
	p_t = \hat{g}^*_{tt}\dot{t} = -E.
\end{equation}
Now using (\ref{massivemaster}) and (\ref{energycon}), we obtain
\begin{equation}
	\dot{r}^2 = -\frac{\hat{g}^*_{tt}+E}{\hat{g}^*_{tt}\hat{g}^*_{rr}}.
\end{equation}
From the above equation, we can calculate the proper time required for a massive particle to reach $ r_* = r_{sing} $ from a finite radius $ r_{in} $. Integrating by parts, we find
\begin{equation}\label{integ1}
	\tau = \int_{r_*}^{r_{in}}\frac{W(r)dr}{\sqrt{W(r)f(r) - E}} \rightarrow \infty.
\end{equation}
The left panel in Figure (\ref{geodesic_a}) shows the numerical integration of the expression (\ref{integ1}). We can see from the plot that it takes infinite amount of time for the massive particle to reach the surface $ r = r_{sing} $. 

Similarly, for a massless particle we have $ \hat{g}^*_{\mu\nu}\dot{x}^\mu\dot{x}^\nu  = 0 $, where the dotted quantities denote dervatives with respect to an affine parameter $ \sigma $. In this case, we find 
\begin{equation}
	\dot{r}^2 = - \frac{E^2}{\hat{g}^*_{tt}\hat{g}^*_{rr}},
\end{equation}
and integrating by parts, we obtain
\begin{equation}\label{integ2}
	\sigma = \int_{r_{sing}}^{r_*}\frac{W(r)dr}{E} \rightarrow \infty.
\end{equation}
The right panel in Figure (\ref{geodesic_a}) shows the numerical integration of the expression (\ref{integ2}). As we can see from the plot, a massless particle cannot reach the singular surface $ r = r_{sing} $ with a finite amount of affine parameter.

\section{Rotating spacetime in massive gravity without complexification}

In this section, we briefly summarize the method without complexification presented by Azreg-Ainou \cite{Azreg-Ainou:2014pra} to construct stationary spacetimes in massive gravity starting from the static metric (5) which can be written as 
\begin{equation}
ds^2=-f(r) dt^2+\frac{dr^2}{g(r)}+h(r)\left(d\theta^2+\sin^2\theta d\phi^2\right).
\end{equation}
The first step of the algorithm is to write down the above metric in the advance null (Eddington-Finkelstein) coordinates $(u,r,\theta,\phi)$ using the transformation
\begin{equation}
du=dt-\frac{dr}{\sqrt{fg}}.
\end{equation}
The metric in the advance null coordinate becomes
\begin{equation}
ds^2=-f(r) du^2-2\sqrt{\frac{f}{g}}dudr+h(r)\left(d\theta^2+\sin^2\theta d\phi^2\right).
\end{equation}
The second step is to express the inverse metric $g^{\mu\nu}$ using a null tetrad $Z_\alpha^\mu=(l^\mu,n^\mu,m^\mu,\bar{m}^\mu)$ in the form
\begin{equation}
g^{\mu\nu}=-l^\mu n^\nu -l^\nu n^\mu +m^\mu \bar{m}^\nu +m^\nu \bar{m}^\mu,
\end{equation}
where $\bar{m}^\mu$ is the complex conjugate of $m^\mu$, and the tetrad vectors satisfy the relations
\begin{equation}
l_\mu l^\mu = n_\mu n^\mu = m_\mu m^\mu = l_\mu m^\mu = n_\mu m^\mu =0,
\end{equation}
\begin{equation}
l_\mu n^\mu = - m_\mu \bar{m}^\mu =-1.
\end{equation}
One finds that the tetrad vectors satisfying the above relations are given by
\begin{equation}
l^\mu=\delta^\mu_r, \, n^\mu=\sqrt{\frac{g}{f}}\delta^\mu_u-\frac{g}{2}\delta^\mu_r, \, m^\mu=\frac{1}{\sqrt{2h}}\left(\delta^\mu_\theta+\frac{i}{\sin\theta}\delta^\mu_\phi\right).
\end{equation}
Then we perform the complex coordinate transformation in the $r-u$ plane given by
\begin{equation}
r\rightarrow r'=r+ia\cos\theta, \hspace{0.3cm} u\rightarrow u'=u-ia\cos\theta,
\end{equation}
where $a$ is the spin parameter. The third step of the Newman–Janis algorithm is usually related with the complexification of the
radial coordinate $r$. Note that there is an ambiguity related to this step, namely, as argued in \cite{Azreg-Ainou:2014pra} there are many ways to complexify $r$, therefore we shall follow here the procedure in Ref. \cite{Azreg-Ainou:2014pra} which basically drops the complexification procedure of the metric functions $f(r)$, $g(r)$ and $h(r)$. In this method, we accept the transformation (24) and that the functions $f(r)$, $g(r)$ and $h(r)$ transform
to $F = F(r, a, \theta)$, $ G = G(r, a, \theta)$ and $H = H(r, a, \theta)$, respectively. Thus our new
null tetrads are 
\begin{equation}
l'^\mu=\delta^\mu_r, \quad n'^\mu=\sqrt{\frac{G}{F}}\delta^\mu_u-\frac{G}{2}\delta^\mu_r,
\end{equation}
\begin{equation}
m'^\mu=\frac{1}{\sqrt{2H}}\left(ia\sin\theta(\delta^\mu_u-\delta^\mu_r)+\delta^\mu_\theta+\frac{i}{\sin\theta}\delta^\mu_\phi\right).
\end{equation}

Using these null tetrads, the new inverse metric given by
\begin{equation}
g^{\mu\nu}=-l'^\mu n'^\nu -l'^\nu n'^\mu +m'^\mu \bar{m}'^\nu +m'^\nu \bar{m}'^\mu.
\end{equation}
The new metric in the Eddington-Finkelstein coordinates reads
\begin{eqnarray}\notag
ds^2&=&-Fdu^2-2\sqrt{\frac{F}{G}}dudr+2a\sin^2\theta\left(F-\sqrt{\frac{F}{G}}\right)du d\phi\\\notag
&+& 2a\sqrt{\frac{F}{G}}\sin^2\theta drd\phi+H d\theta^2 \\
&+&\sin^2\theta\left[H+a^2\sin^2\theta\left(2\sqrt{\frac{F}{G}}-F\right)\right]d\phi^2.
\label{eq:null_coordinate_metric_1}
\end{eqnarray}
The final but crucial step is to bring this form of the metric to the Boyer-Lindquist
coordinates by a global coordinate transformation of the form
\begin{equation}
du=dt'+\va(r)dr, \hspace{0.5cm} d\phi=d\phi'+\chi(r) dr.
\label{eq:transformation_to_BL}
\end{equation}
Here
\begin{equation}
\va(r)=-\frac{k(r)+a^2}{g(r)h(r)+a^2},
\end{equation}
\begin{equation}
\chi(r)=-\frac{a}{g(r)h(r)+a^2},
\end{equation}
\begin{equation}
k(r)=\sqrt{\frac{g(r)}{f(r)}}h(r).
\end{equation}
Since the functions $F$, $G$ and $H$ are still unknown, one can fix some of them to get
rid of the cross-term $dtdr$ in the metric. Now, if we choose
\begin{equation}
F(r)=\frac{(g(r) h(r)+a^2 \cos^2\theta) H}{(k(r)+a^2 \cos^2\theta)^2},
\end{equation}
\begin{equation}
G(r)=\frac{g(r) h(r)+a^2 \cos^2\theta}{H}.
\end{equation}
The rotating solution is finally written as
\begin{equation}\notag
ds^2 = -\frac{(g(r) h(r)+a^2 \cos^2\theta) H}{(k(r)+a^2 \cos^2\theta)^2}dt^2+\frac{H dr^2}{g(r) h(r)+a^2}
\end{equation}
\begin{equation}
- 2 a \sin^2\theta \left[ \frac{k(r)-g(r)h(r)}{(k(r)+a^2 \cos^2\theta)^2}\right]H dt d\phi+H d\theta^2\\\notag
\end{equation}
\begin{equation}
+H \sin^2\theta \Big[ 1+a^2 \sin^2\theta \Big(\frac{2k(r)-g(r)h(r)+a^2 \cos^2\theta}{(k(r)+a^2 \cos^2\theta)^2}\Big) \Big] d\phi^2.
\end{equation}

Now since $f(r)=g(r)$,  and $h(r)=r^2$, thus Eq. (33) implies $k(r) = h(r)$. Furthermore, the function $H(r, \theta, a)$ is still arbitrary and can be chosen so that the
cross-term of the Einstein tensor $G_{r \theta}$, for a physically acceptable rotating solution,
identically vanishes, i.e. $G_{r \theta}=0$. The latter constraint yields the differential equation 
\begin{eqnarray}
(h(r)+a^2 y^2)^2(3 H_{,r}H_{,y^2}-2H H_{,r y^2})=3 a^2 h_{,r}H^2,
\end{eqnarray}
where $y=\cos \theta$ and $h(r)=r^2$. One can check that the solution of the above equation has the following form (see, \cite{Azreg-Ainou:2014pra})
\begin{eqnarray}
H=h(r)+a^2\cos^2\theta=r^2+a^2 \cos^2\theta.
\end{eqnarray}

\begin{figure*}[t]
	\begin{center}
		\includegraphics[type=pdf,ext=.pdf,read=.pdf,width=7.0cm]{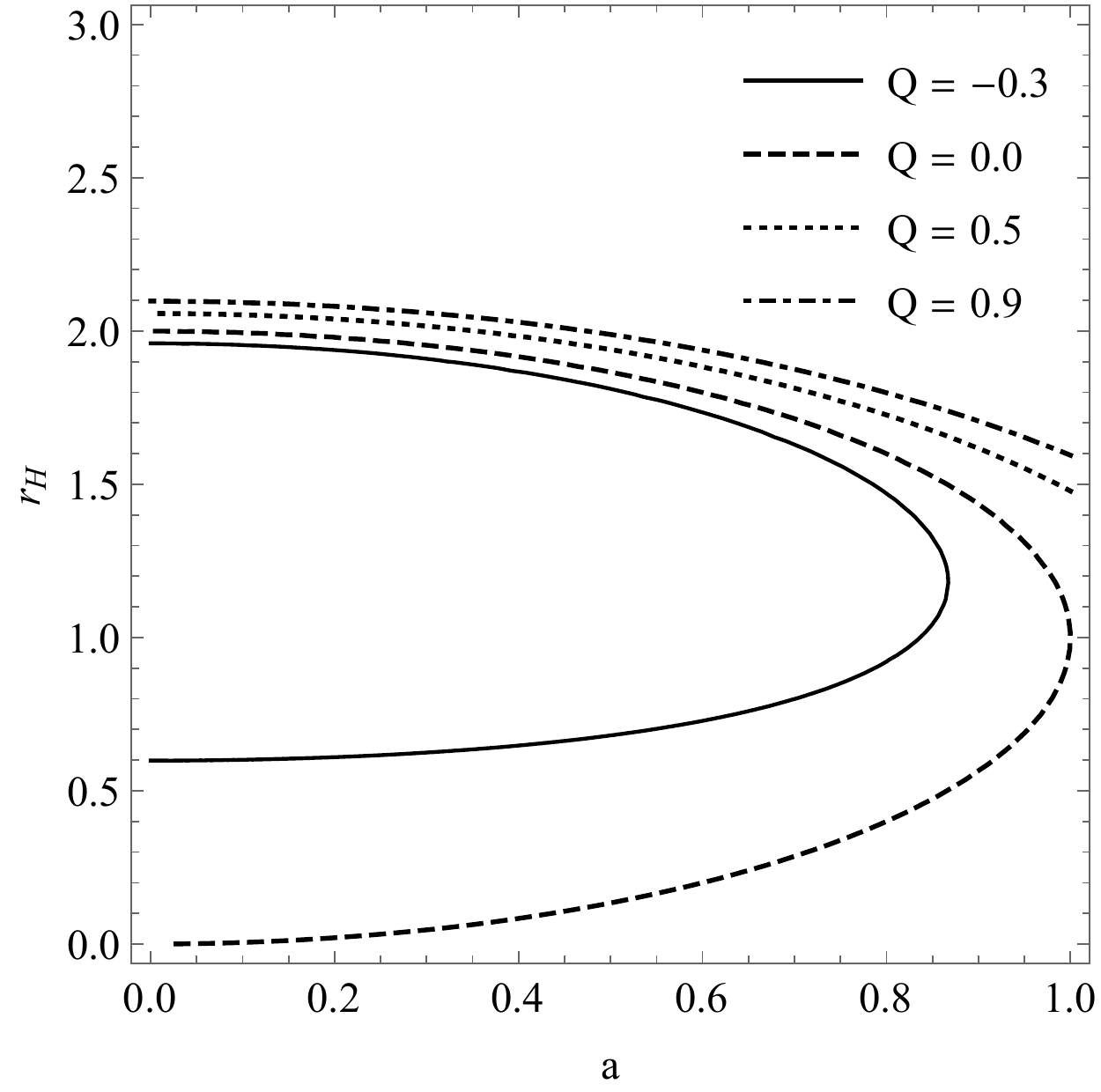}
		\includegraphics[type=pdf,ext=.pdf,read=.pdf,width=7.0cm]{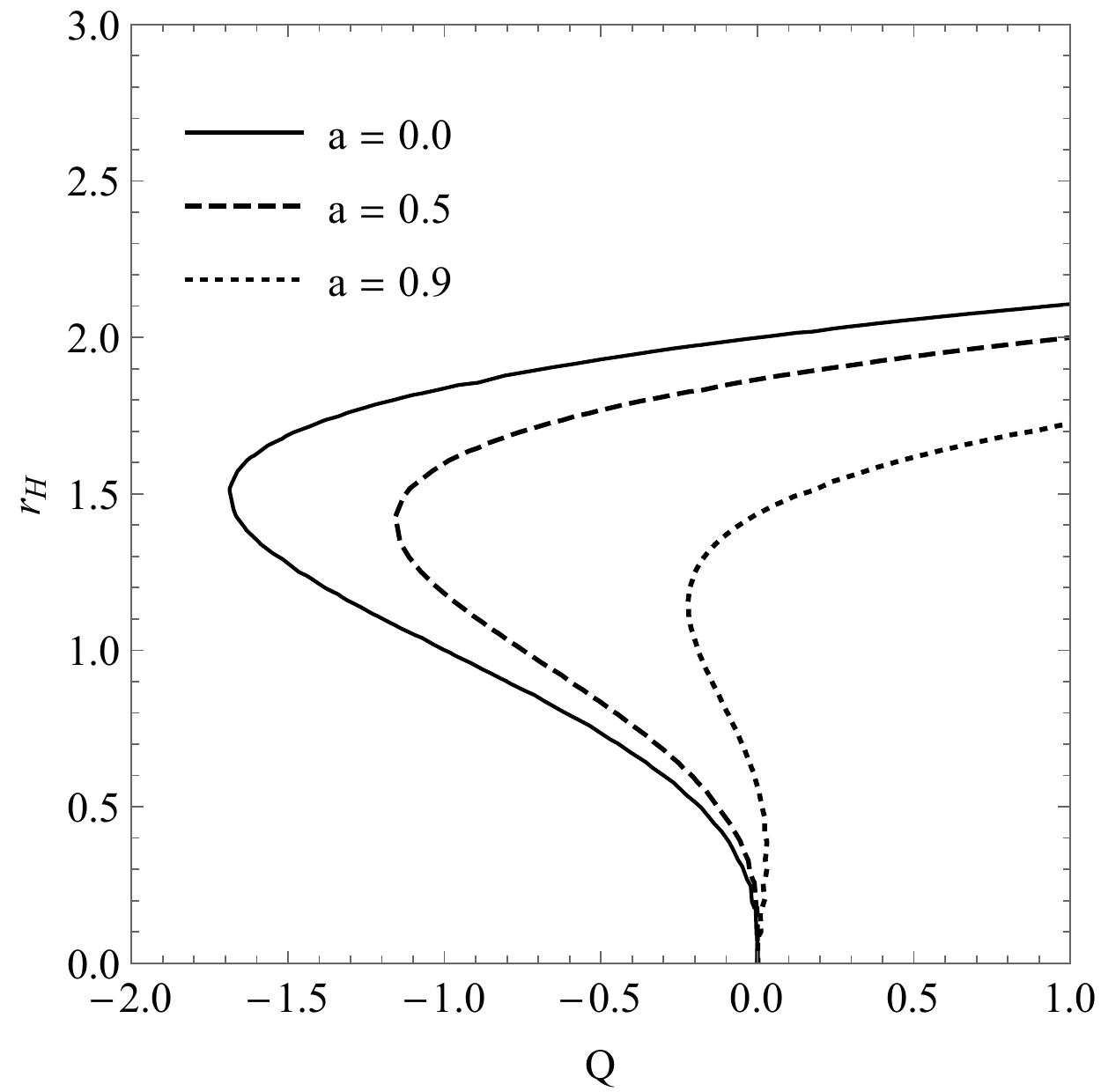}
	\end{center}
	\vspace{-0.5cm}
	\caption{Left panel: Radius of horizon $ r_H $ plotted against the spin $ a $ for $ Q = -0.3, 0.0, 0.5 \text{ and } 0.9 $. For  $ Q \leq 0 $, we see both outer and inner horizons except for the extremal case. For $ Q > 0 $, only outer horizon is present. Right panel: Radius of horizon $ r_H $ plotted against the scalar charge $ Q $ for $ a = 0.0, 0.5 \text{ and } 0.9 $. We see similar behavior as the left panel, where there are two horizons for $ Q \leq 0 $, but only a single horizon for $ Q > 0 $ for different values of the spin parameter $ a $. Here $ M=1 $ and $ \lambda = 4 $. \label{horizon}}
\end{figure*}

With this information in hand, the rotating black hole solution in massive gravity reads
\begin{equation}\notag
ds^2 = -\left(1-\frac{2Mr+Q r^{2-\lambda}}{\rho^2}\right)dt^2+\frac{\rho^2 dr^2}{\Delta}
\end{equation}
\begin{equation}
- 2 a \sin^2\theta \left[\frac{2Mr+Qr^{2-\lambda}}{\rho^2}\right] dt d\phi+\rho^2 d\theta^2\\\notag
\end{equation}
\begin{equation}\label{yyu}
+\sin^2\theta \Big[ r^2+a^2+a^2\sin^2\theta\left(  \frac{2Mr+Qr^{2-\lambda}}{\rho^2}\right) \Big] d\phi^2,
\end{equation}
where 
\begin{equation}
    \rho^2=r^2+a^2\cos^2\theta,
\end{equation}
and
\begin{equation}\label{delta}
    \Delta=g(r)h(r)+a^2=r^2-2Mr-Qr^{2-\lambda}+a^2.
\end{equation}
This spacetime is singular at the surface $ r = r_{sing} $, where $ \rho^2 = 0 $. We plot the radius of horizons with respect to the spin $ a $ and scalar charge $ Q $ in Fig. (\ref{horizon}).   For instance, Fig. (\ref{horizon}) (left panel) shows that there are exactly two horizons for each value of spin parameter except near the turning points where a unique horizon exists (extremal case). Fig. (\ref{horizon}) (right) shows the existence of two horizons explicitly when $Q<0$. For $Q>0$, a single horizon exists (suggesting the extremal case).

\section{Rotating regular black holes in conformal massive gravity}

In this section, we construct singularity-free rotating black holes. In particular, the singularity-free black hole solution is obtained by rescaling the Kerr metric in massive gravity by a suitable overall conformal factor $W(r,\theta)$ as follows
\begin{equation}\label{NRBH} 
ds^{* 2} \equiv 
\hat{g}_{\mu\nu}^* dx^\mu dx^\nu = W(r,\theta) \hat{g}_{\mu\nu} dx^\mu dx^\nu, 
\end{equation}
where
\begin{equation}
W(r,\theta) =\left( 1 + \frac{L^2}{r^2+a^2\cos^2\theta} \right)^{\frac{|\lambda|}{2}+2}.
\end{equation}

\begin{figure*}[t]
	\begin{center}
		\includegraphics[type=pdf,ext=.pdf,read=.pdf,width=7.0cm]{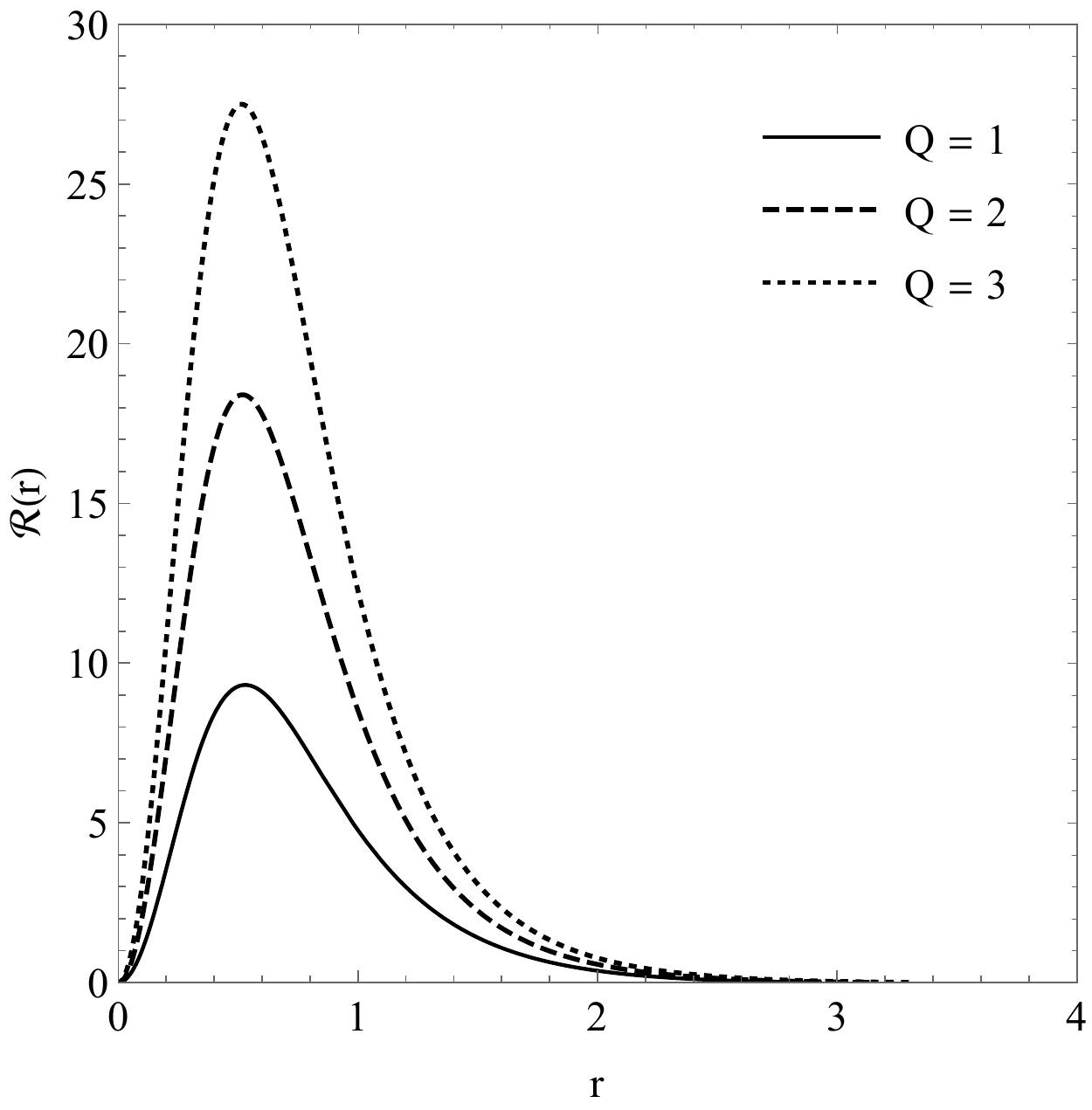}
		\includegraphics[type=pdf,ext=.pdf,read=.pdf,width=7.0cm]{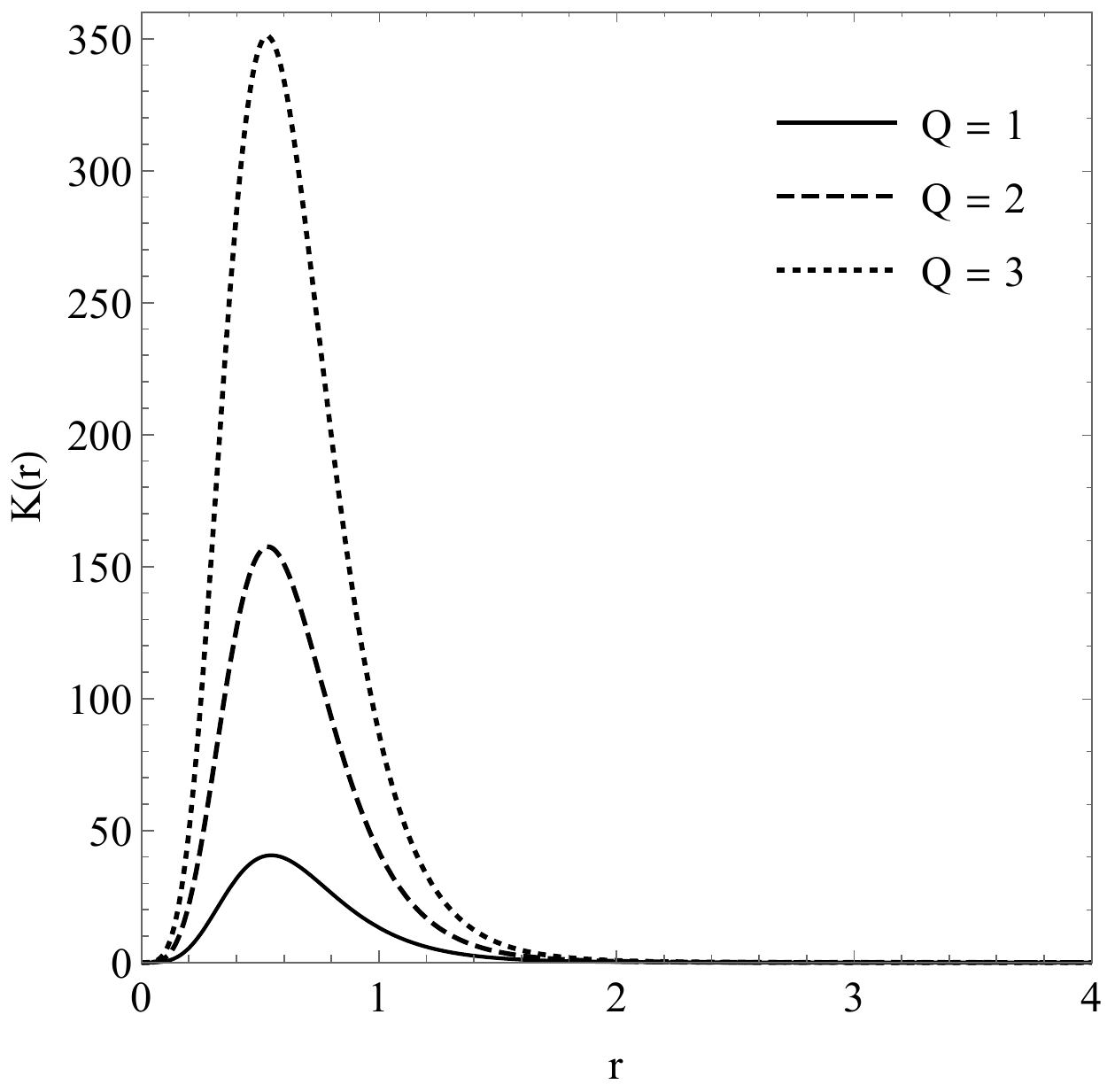}
	\end{center}
	\vspace{-0.5cm}
	\caption{Left panel: Ricci scalar as a function of the radial coordinate $ r $. Right panel: Kretschmann invariant as a function of radial coordinate $ r $. We can see in both plots, a higher curvature region close to the black hole, but is finite at the location of the singularity. In both plots, the solid line, dashed line and dotted line corresponds to $ Q = 1 $, $ Q = 2 $ and $ Q = 3 $, respectively. We assume $M =L=1$, $ a = 0.7 $, $ \lambda = 4 $ and $ \theta = \pi/2 $. \label{curvature}}
\end{figure*}

In general, one can have an infinite class of such functions $W(r)$ that enable us to map the singular Schwarzschild spacetime to an "everywhere regular" one. The metric after conformal rescaling can be written as
\begin{eqnarray}
ds^{*2} &=&  \left(1 + \frac{L^2}{\rho^2}\right)^{\frac{|\lambda|}{2}+2} ds^2,  
\end{eqnarray}
where $ds^2$ is given by Eq. (\ref{yyu}). For the metric above, the expressions for the Kretschmann invariant, ${ \hat{K} } = {\bf \hat{R}iem}^2$  and Ricci scalar are cumbersome. We show their behavior with respect to radial coordinate on the equatorial plane in Fig. (\ref{curvature}), which suggests a presence of highly curved/non-flat spacetime regions $0<r<2$, while it is nearly flat outside this domain. It is interesting to note that there are no spacetime divergences.

\begin{figure*}[t]
	\begin{center}
		\includegraphics[type=pdf,ext=.pdf,read=.pdf,width=7.0cm]{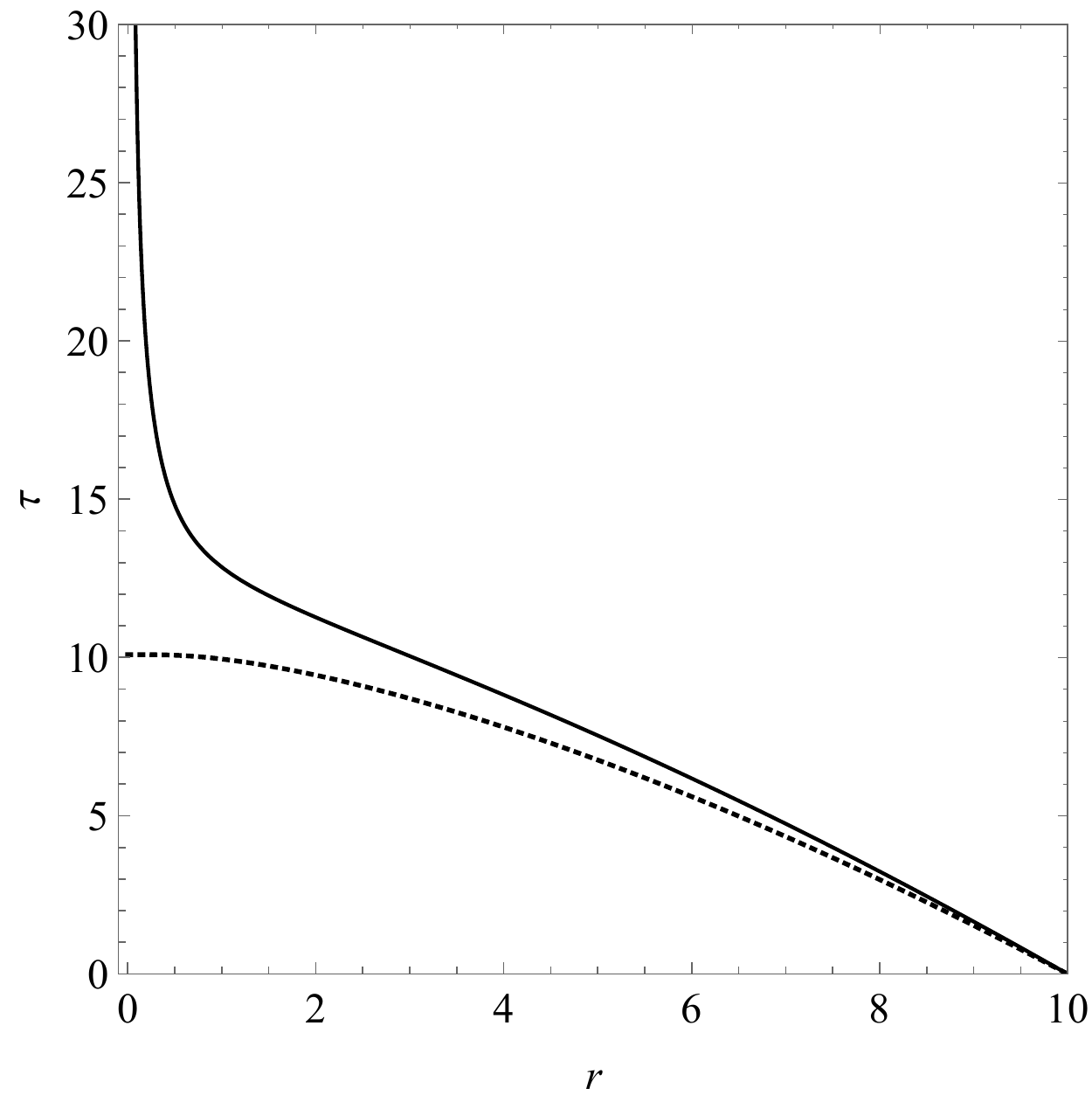}
		\includegraphics[type=pdf,ext=.pdf,read=.pdf,width=7.0cm]{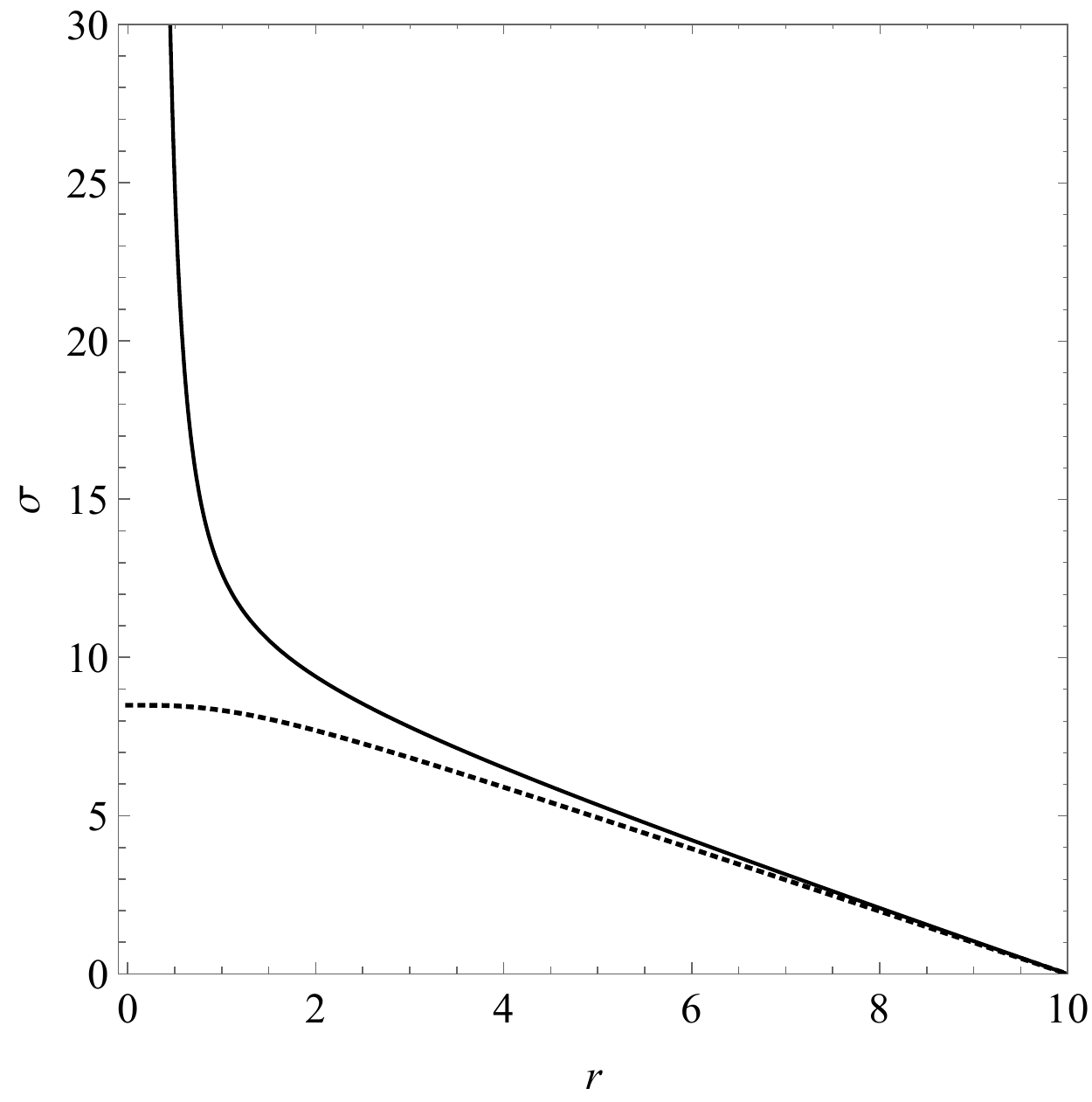}
	\end{center}
	\vspace{-0.5cm}
	\caption{Left panel: proper time $\tau$ as a function of the radial coordinate $r$ of a massive particle with vanishing angular momentum moving to smaller radii from the initial coordinate $r = r_{\rm in}$. The solid line corresponds to the proper time in the rescaled metric and the particle cannot reach the surface $r = r_{\rm sing}$. The dashed line corresponds to the standard metric in massive gravity. Right panel: as in the left panel for the affine parameter $\sigma$ of a massless particle. In these plots, we assume $E=L=1$, $ M = 2 $, $ \lambda = 4 $, $ Q = 0.5 $, $ \theta = \pi/2 $, and $r_{\rm in}=10$. \label{geodesic_b}}
\end{figure*}

\begin{figure*}
\includegraphics[width=7.1cm]{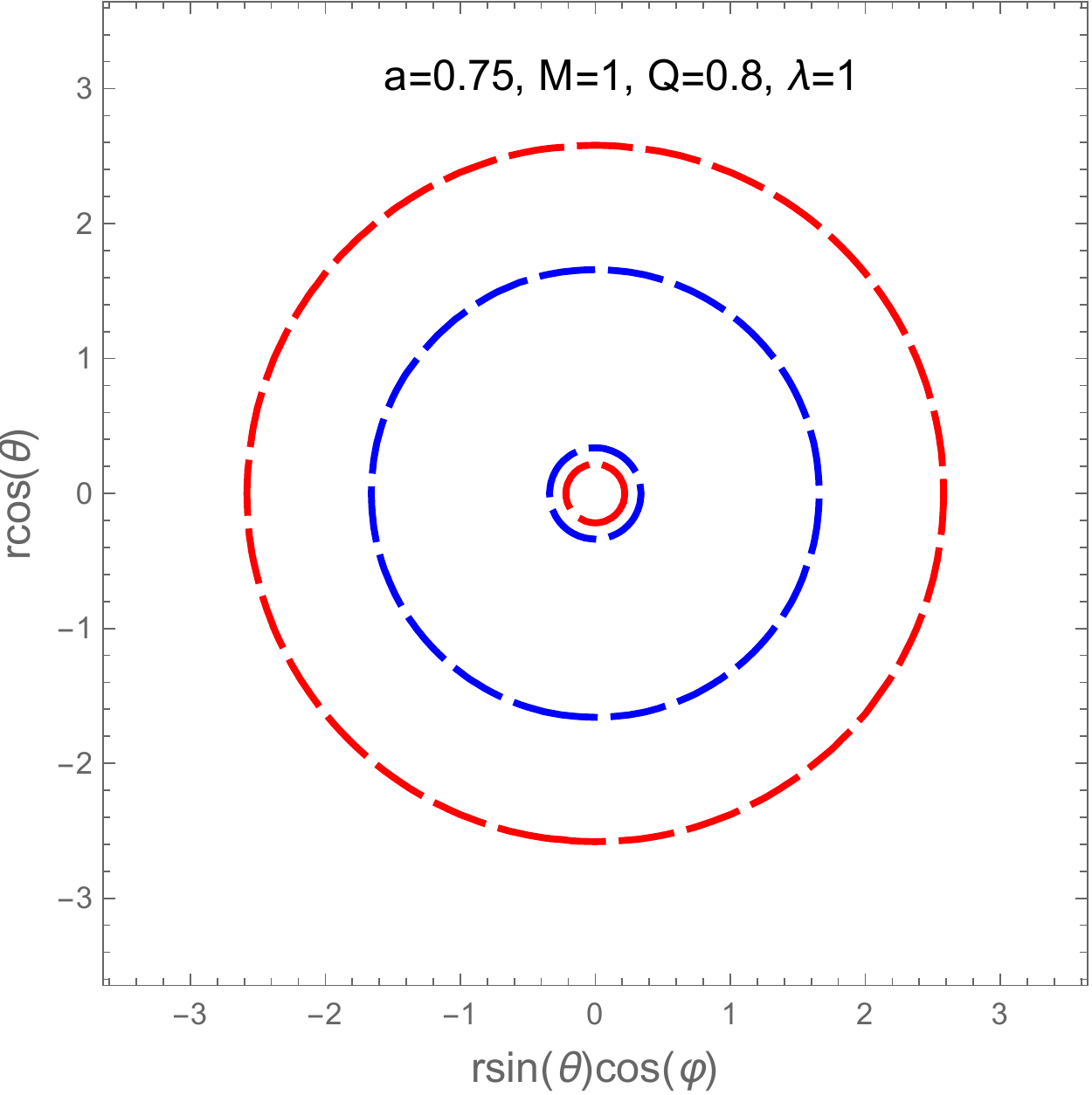}
\includegraphics[width=7.1cm]{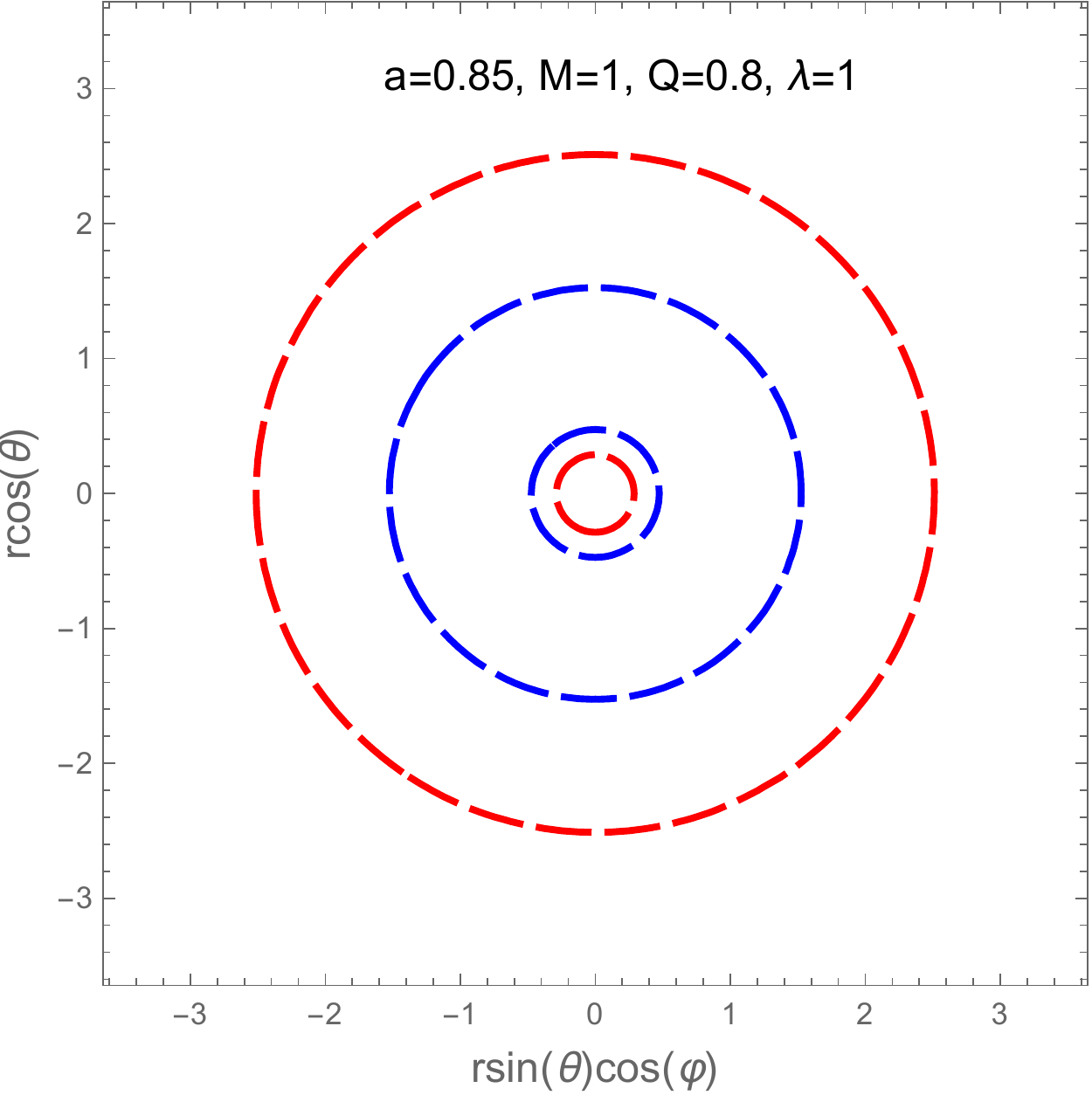}
\caption{The size of the event horizon of the black hole in massive gravity (red color) compared to the Kerr vacuum black hole ( blue color with $Q=0$). We observe that for positive $Q$ and constant $\lambda$, the size of the event horizon is bigger compared to the Kerr vacuum case.} 
\end{figure*}

\begin{figure*}
\includegraphics[width=7.1cm]{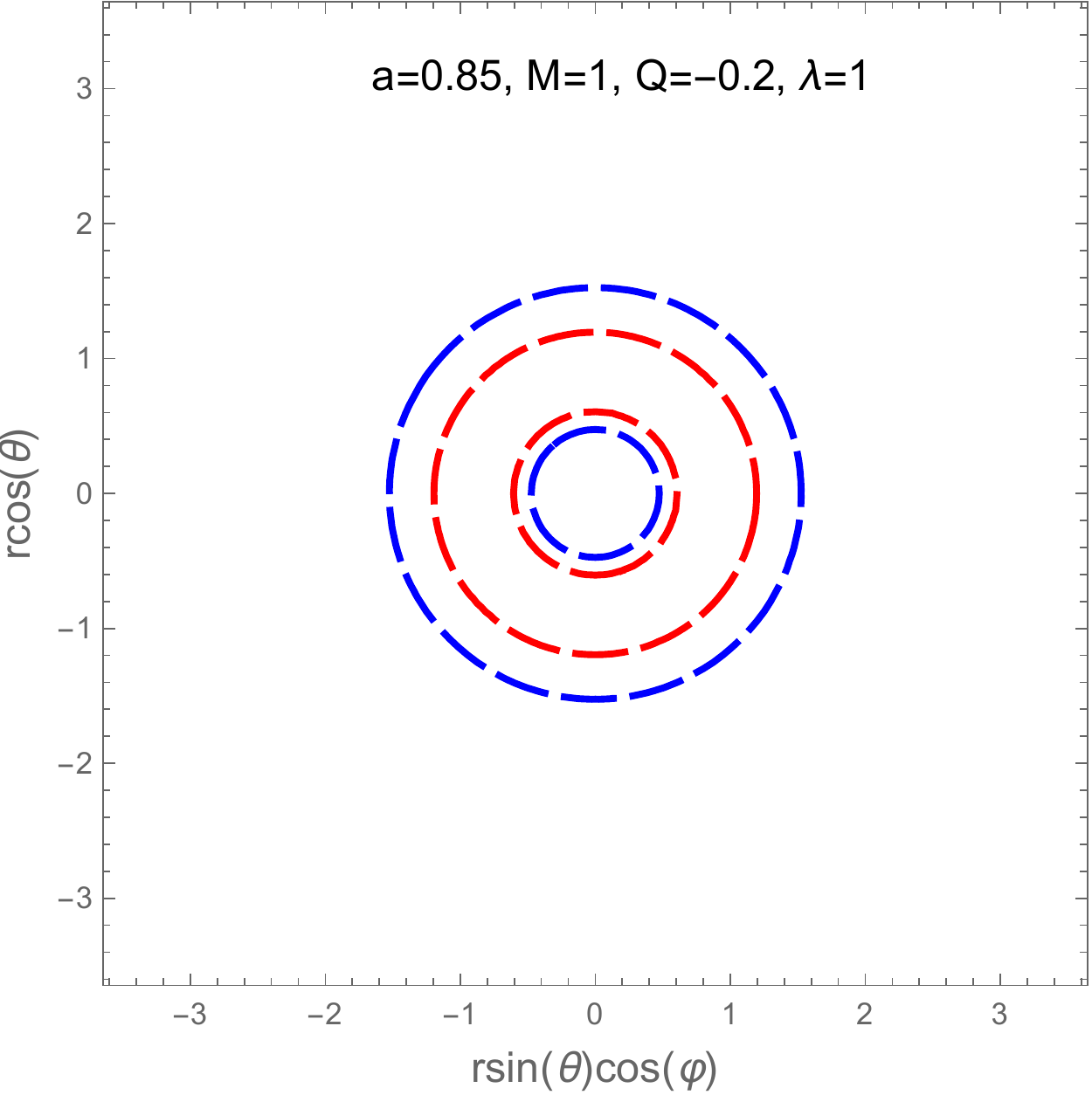}
\includegraphics[width=7.1cm]{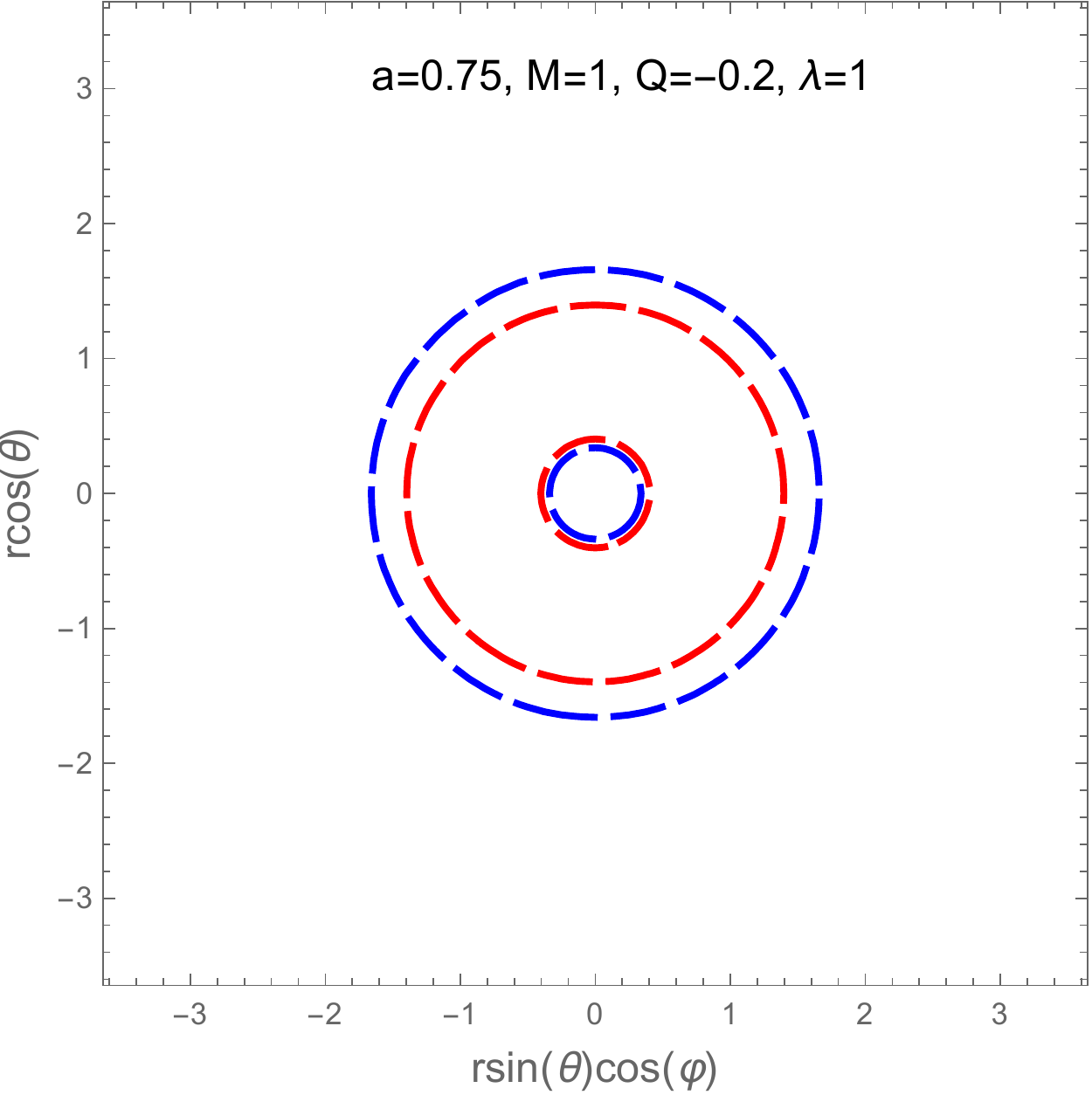}
\caption{The size of the event horizon of the black hole in massive gravity (red color) compared to the Kerr vacuum black hole ( blue color with $Q=0$). We see that for negative values of $Q$ and constant $\lambda$, the event horizon size  is smaller compared to the Kerr vacuum case. } \label{B}
\end{figure*}

Now we show the regularity of the spacetime by studying the geodesic completion of massive and massless particles. We start with the Lagrangian for a massive particle moving in the equatorial plane $ (\theta = \pi/2) $
\begin{equation}
    \begin{aligned}
        \mathcal{L}_m & = -m\sqrt{-\dot x^\mu \dot x_\mu} \\
        & = -m\sqrt{-(\hat{g}^*_{tt}\dot{t}^2+\hat{g}^*_{rr}\dot{r}^2+2\hat{g}^*_{t\phi}\dot{t}\dot{\phi}+\hat{g}^*_{\phi\phi}\dot{\phi}^2)},
    \end{aligned}
\end{equation}
where $ m $ is the mass of the test particle. The metric here is stationary and axis-symmetric, therefore we shall have two invariant quantities given by
\begin{equation}\label{inv2}
\begin{aligned}
& E = \frac{m^2}{\mathcal{L}_m}(\hat{g}^*_{tt}\dot{t}+\hat{g}^*_{t\phi}\dot{\phi}), \\
& J = -\frac{m^2}{\mathcal{L}_m}(\hat{g}^*_{t\phi}\dot{t}+\hat{g}^*_{\phi\phi}\dot{\phi}).
\end{aligned}
\end{equation}
Solving the above equations for $ \dot{\phi} $ and $ \dot{t} $ with $ J = 0 $ and $ \mathcal{L} = -m $, we obtain
\begin{equation}\label{phidots}
\begin{aligned}
& \dot{\phi} = - \frac{\hat{g}^*_{t\phi}}{\hat{g}^*_{\phi\phi}}\dot{t}, \\
& \dot{t} = -\frac{E}{m}\left( \frac{\hat{g}^*_{\phi\phi}}{\hat{g}^*_{tt}\hat{g}^*_{\phi\phi} - \hat{g}^{*2}_{t\phi}} \right).
\end{aligned}  
\end{equation}
The equation of motion for a massive particle is 
\begin{equation}\label{eom2}
\hat{g}^*_{tt}\dot{t}^2+\hat{g}^*_{rr}\dot{r}^2+2\hat{g}^*_{t\phi}\dot{t}\dot{\phi}+\hat{g}^*_{\phi\phi}\dot{\phi}^2 = -1.
\end{equation} 
Using (\ref{phidots}), we write (\ref{eom2}) as
\begin{equation}
\hat{g}^*_{rr}\dot{r}^2 + \mathcal{E}^2\left(  \frac{\hat{g}^*_{\phi\phi}}{\hat{g}^*_{tt}\hat{g}^*_{\phi\phi} - \hat{g}^{*2}_{t\phi}}  \right) = -1,
\end{equation}
where $ \mathcal{E} = E/m $ as before. Rearranging this differential equation, we can calculate the proper time required for a massive particles to reach the surface $ r = r_{sing} $,
\begin{equation}
\tau = -\int_{r_*}^{r_{in}} \frac{dr}{\sqrt{\frac{-1}{\hat{g}^*_{rr}}\left( 1 + e^2\left(  \frac{\hat{g}^*_{\phi\phi}}{\hat{g}^*_{tt}\hat{g}^*_{\phi\phi} - \hat{g}^{*2}_{t\phi}}  \right) \right)}}.
\end{equation}  
We numerically solve this integral and the result is shown in the left panel of Figure (\ref{geodesic_b}). One can easily see that the massive particle requires infinite amount of proper time to reach the surface $ r = r_{sing} $ in the non-singular rotating metric in massive gravity. On the other hand, a massive particle in the unscaled metric reaches the surface $ r = r_{sing} $ at a finite time.  
Similarly for massless particles, we have $ \hat{g}_{\mu\nu}^*\dot{x}^{\mu}\dot{x}^{\nu} = 0 $. Here the dotted quantities are now derivatives with respect to an affine parameter $ \sigma $. The equation of motion for a massless particle with vanishing angular momentum becomes
\begin{equation}
\hat{g}^*_{rr}\dot{r}^2 + \frac{e^2\hat{g}^*_{\phi\phi}}{\hat{g}^*_{tt}\hat{g}^*_{\phi\phi} - \hat{g}^{*2}_{t\phi}} = 0.
\end{equation}
We can integrate this differential equation for a photon directed towards $ r = r_{sing} $, and the result reads
\begin{equation}
\sigma = \int_{r_*}^{r_{in}}\frac{dr}{\sqrt{\frac{-e^2\hat{g}^*_{\phi\phi}}{\left(\hat{g}^*_{tt}\hat{g}^*_{\phi\phi} - \hat{g}^{*2}_{t\phi}\right)\hat{g}^*_{rr}}}}.
\end{equation}
We plot the solution of this integral in the right panel of Figure (\ref{geodesic_b}). It is clear from the plot that the massless particle never reaches the surface $ r = r_{sing} $ with finite amount of affine parameter.

\begin{figure*}
\includegraphics[width=7.1cm]{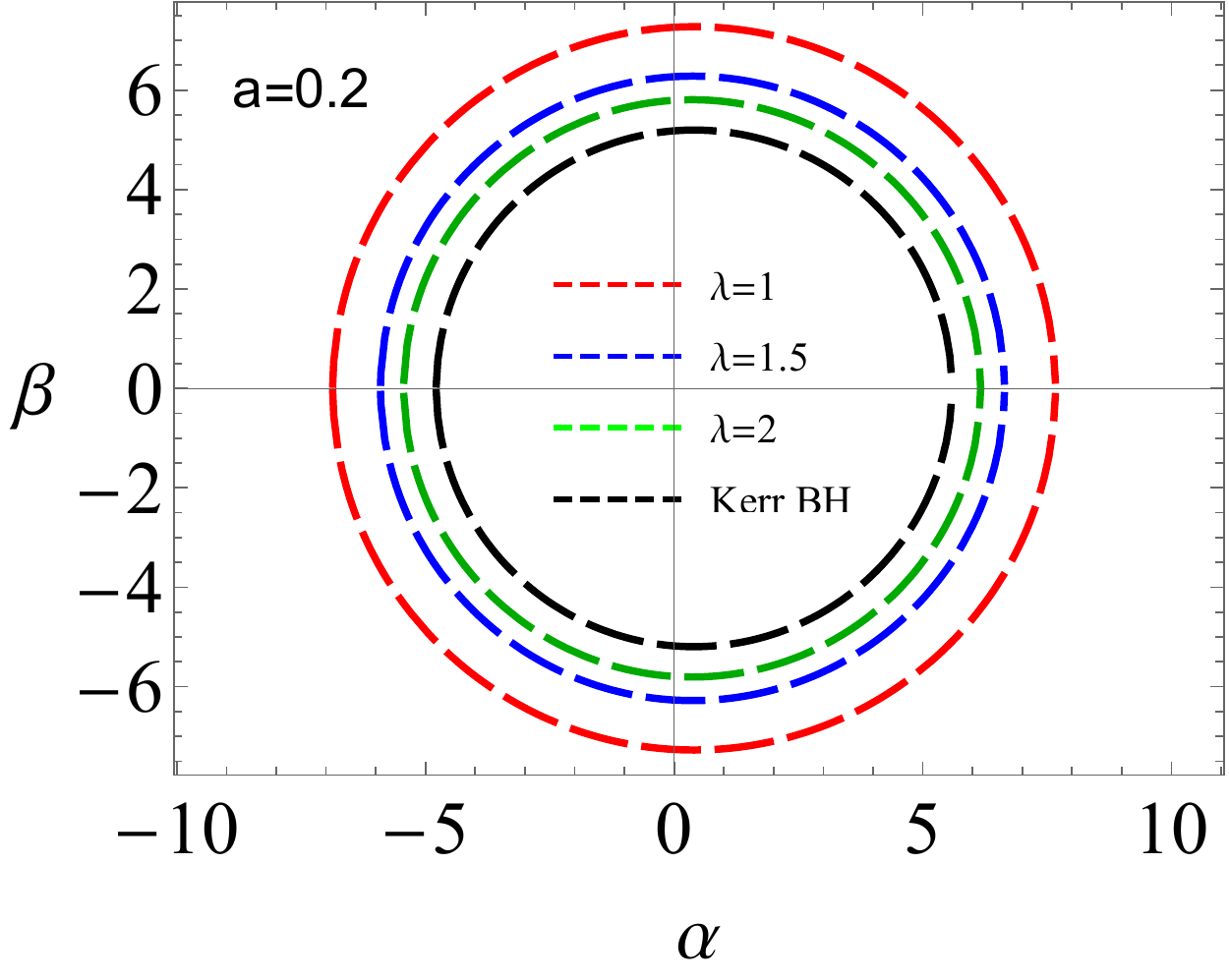}
\includegraphics[width=7.1cm]{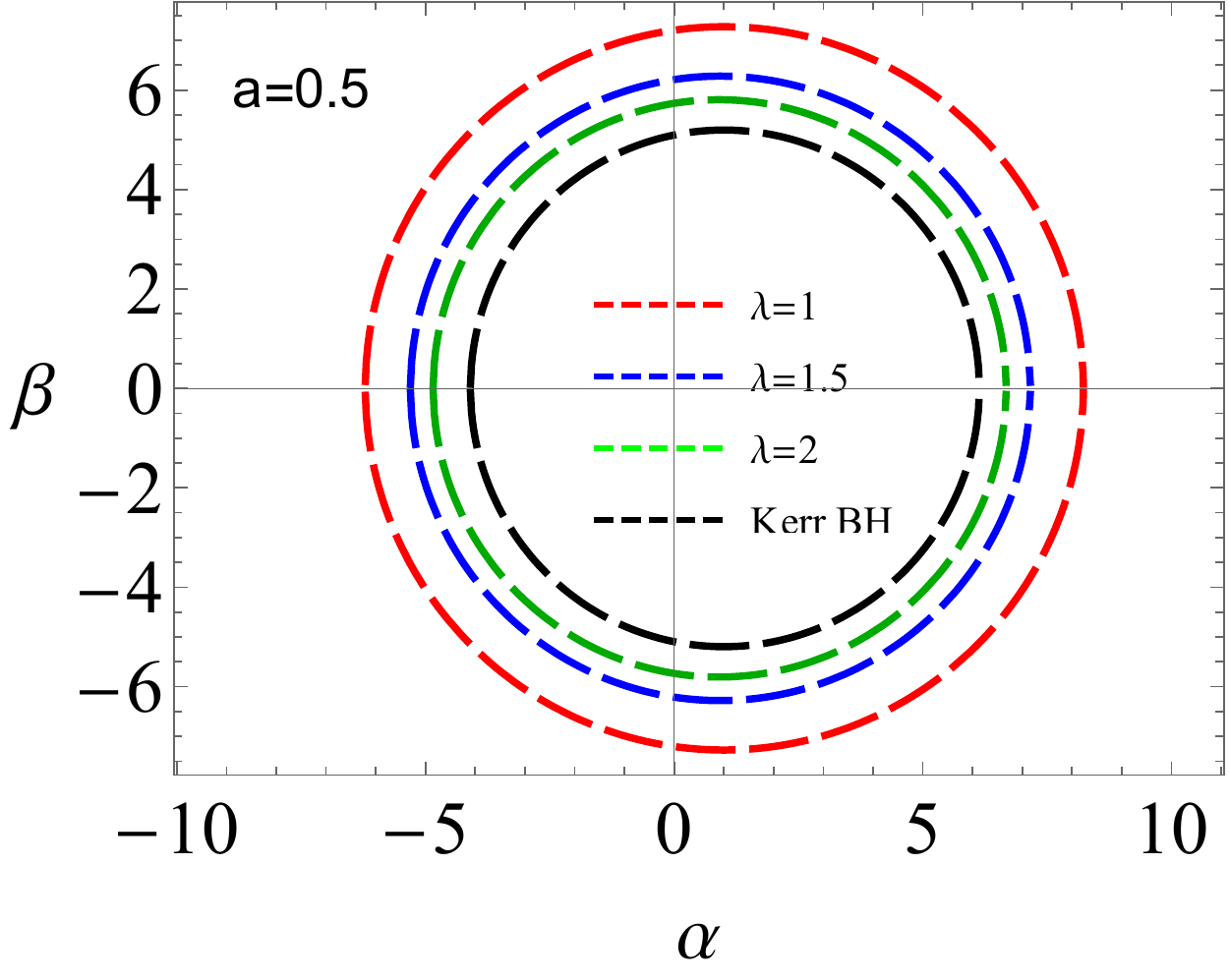}
\includegraphics[width=7.1cm]{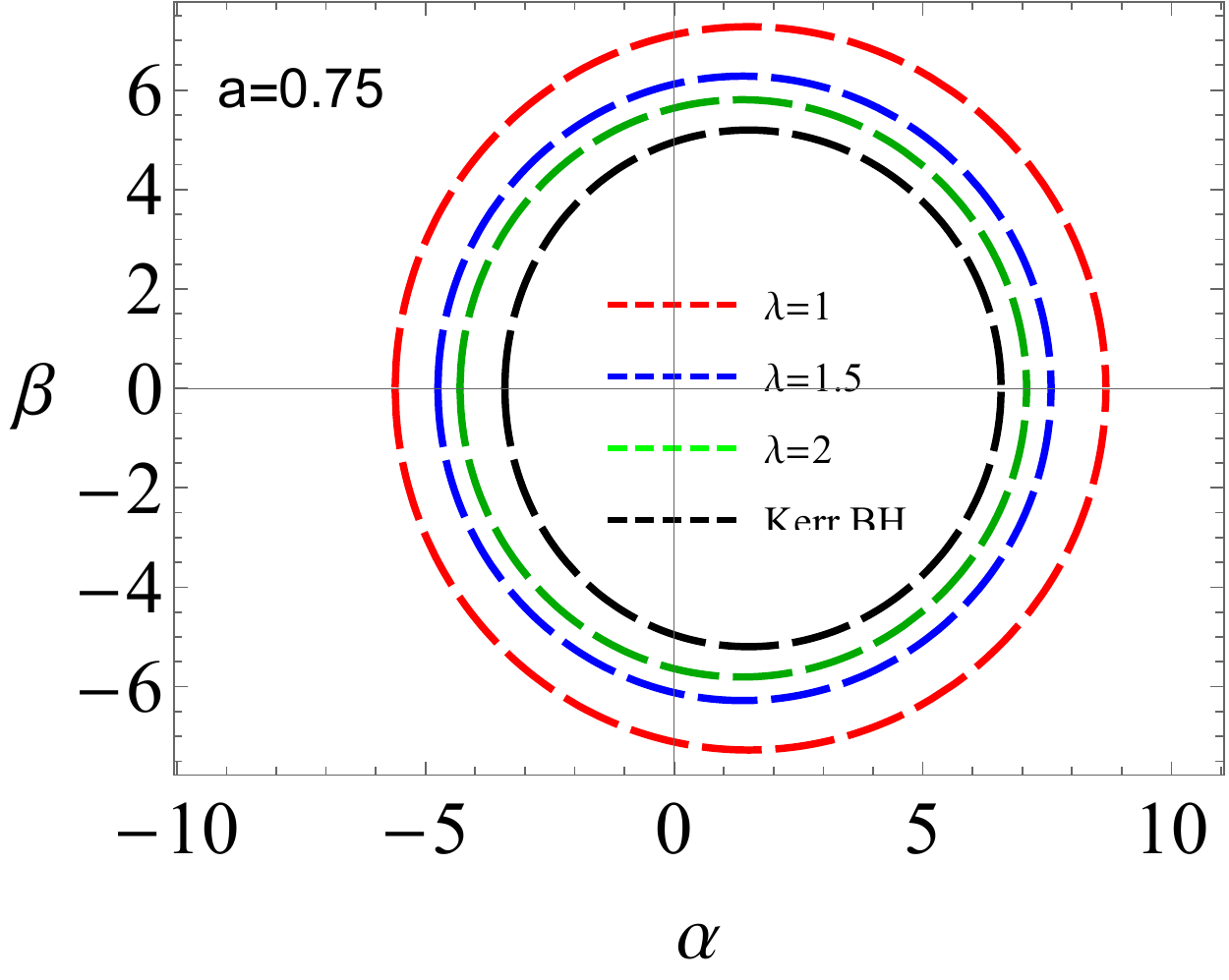}
\includegraphics[width=7.1cm]{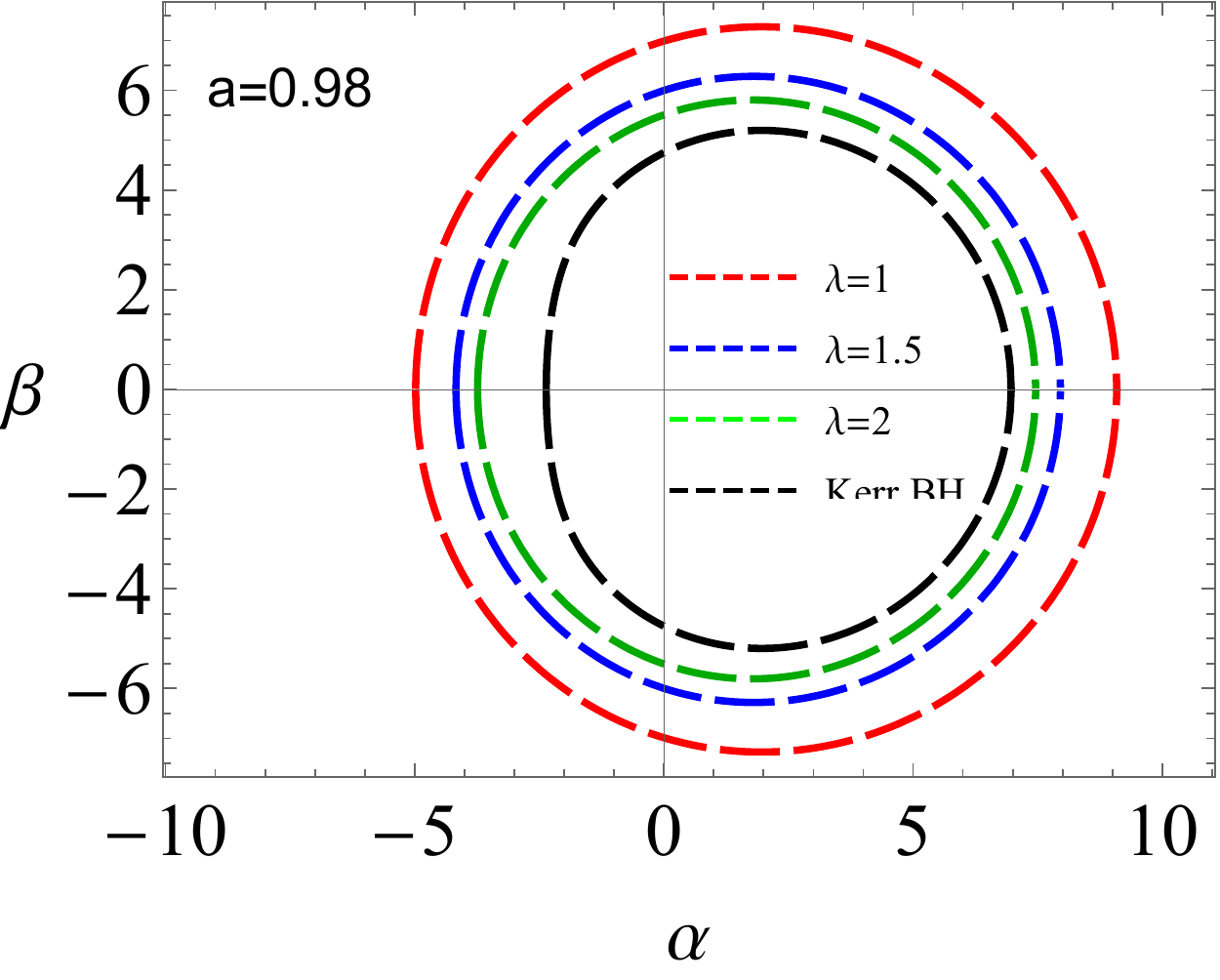}
\caption{The shape of shadow for $M=1$ and $Q=0.8$ and positive $\lambda$.  } \label{B}
\end{figure*}


\begin{figure*}
\includegraphics[width=7.1cm]{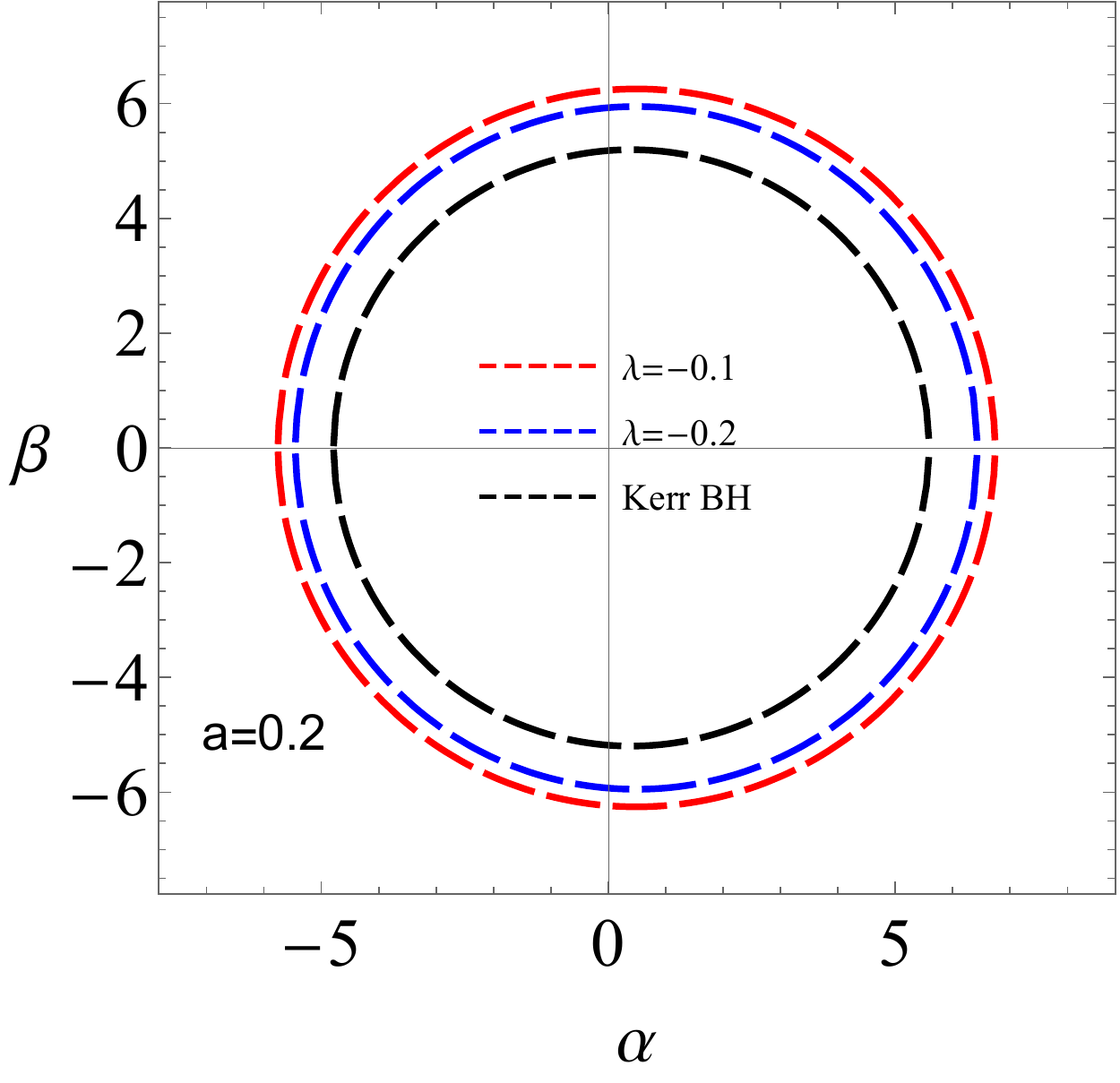}
\includegraphics[width=7.1cm]{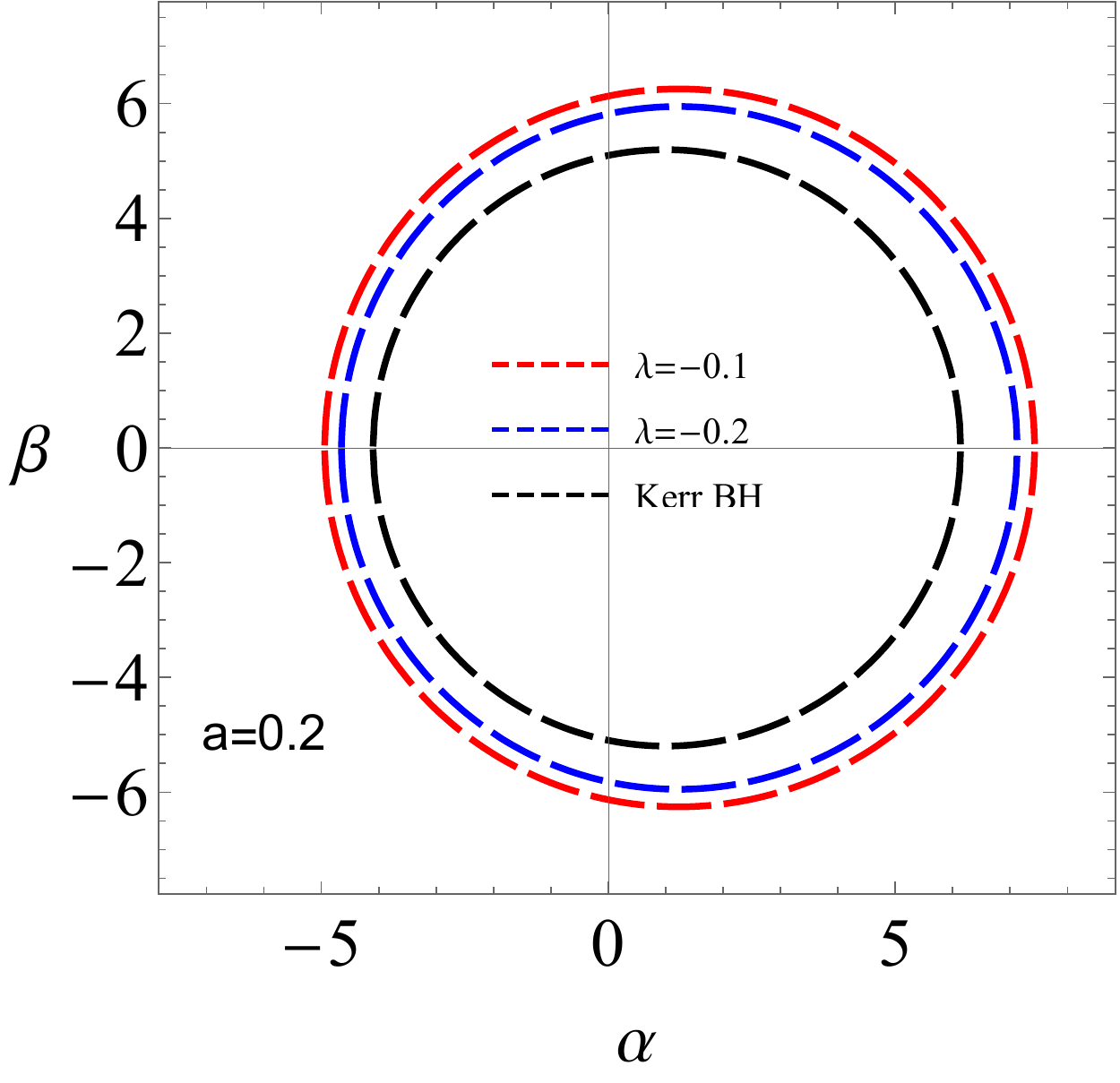}
\includegraphics[width=7.1cm]{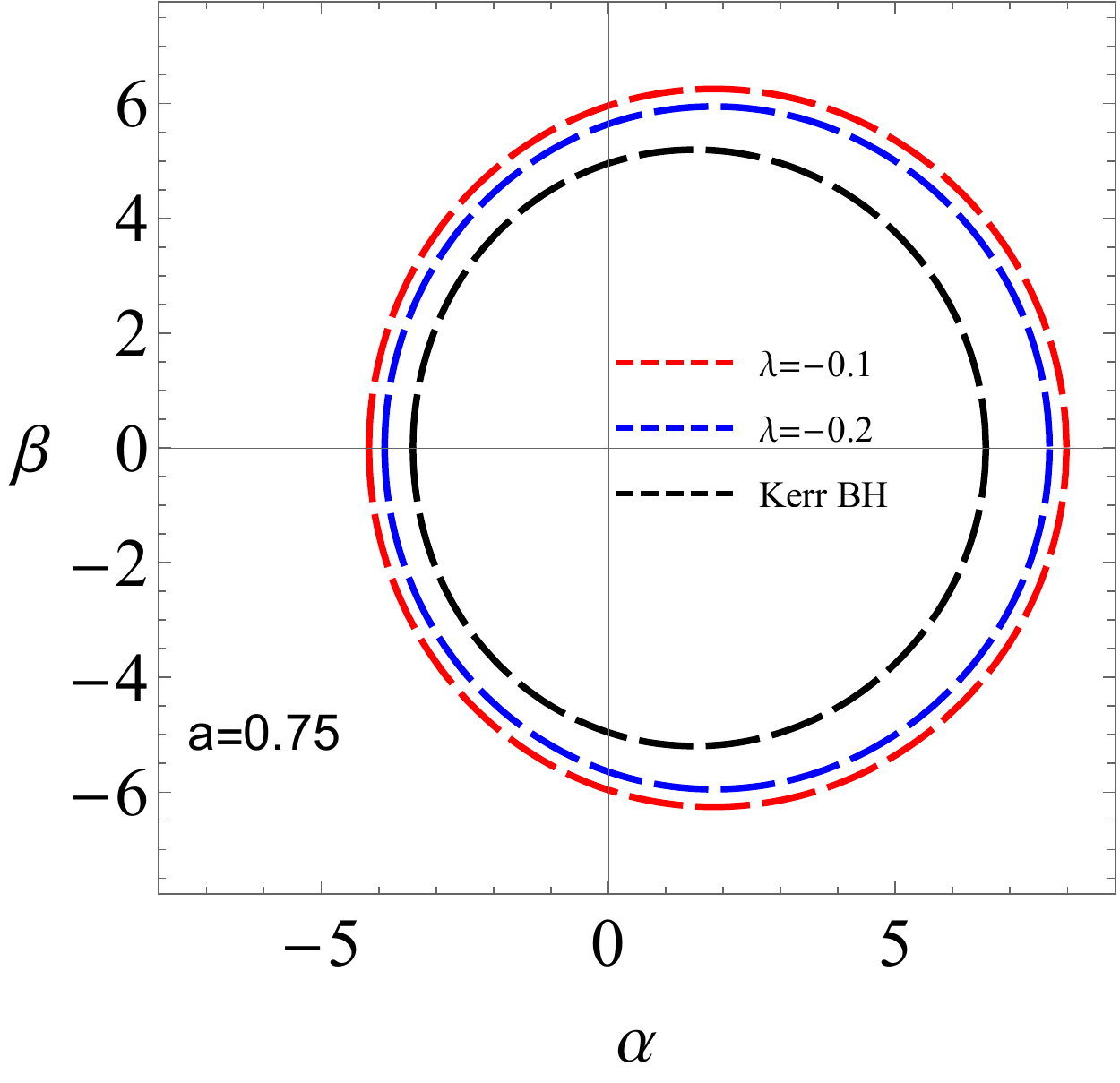}
\includegraphics[width=7.1cm]{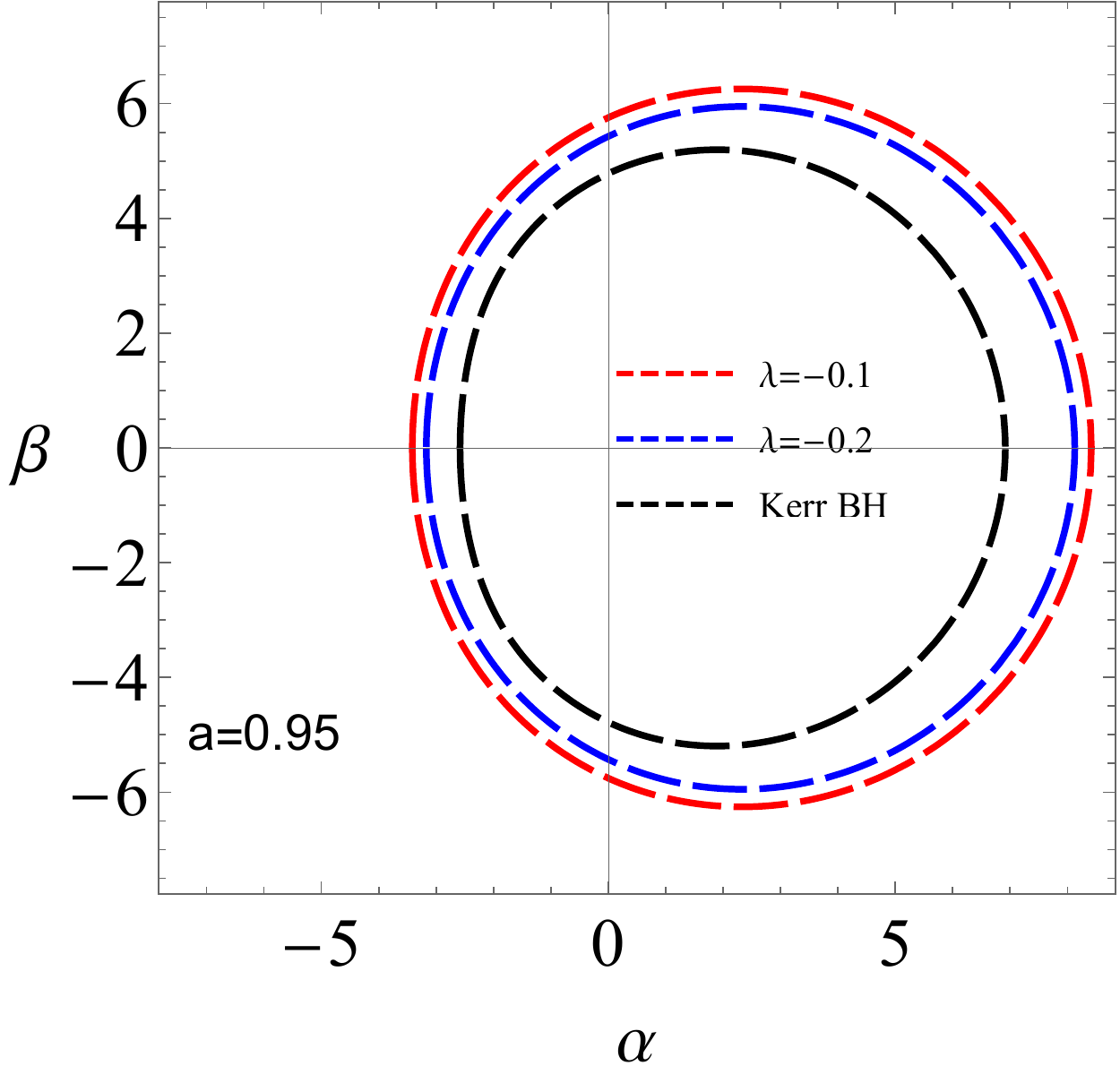}
\caption{The shape of shadow for $M=1$ and $Q=0.2$ and negative $\lambda$.  } \label{B}
\end{figure*}

\begin{figure*}
\includegraphics[width=7.1cm]{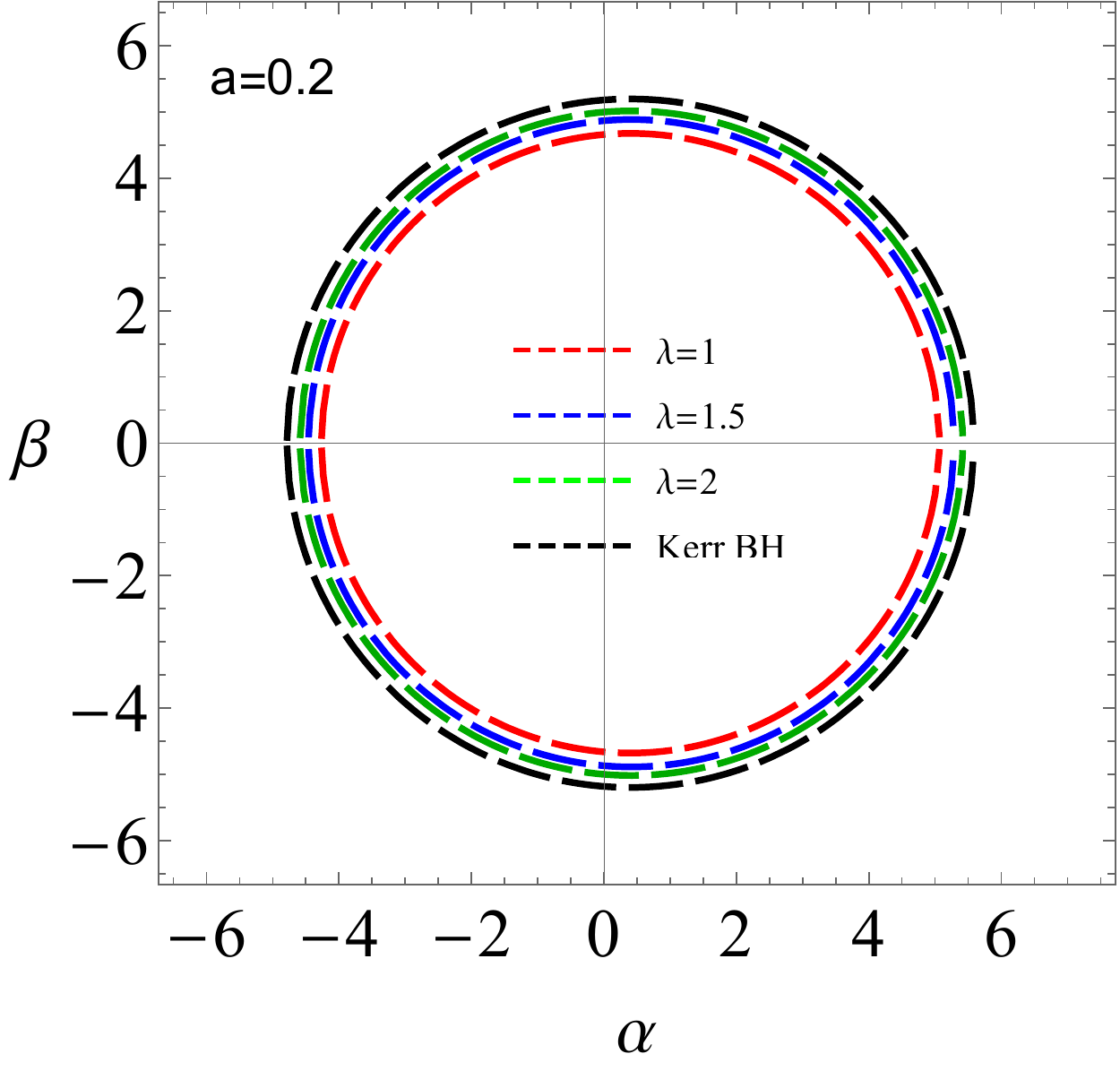}
\includegraphics[width=7.1cm]{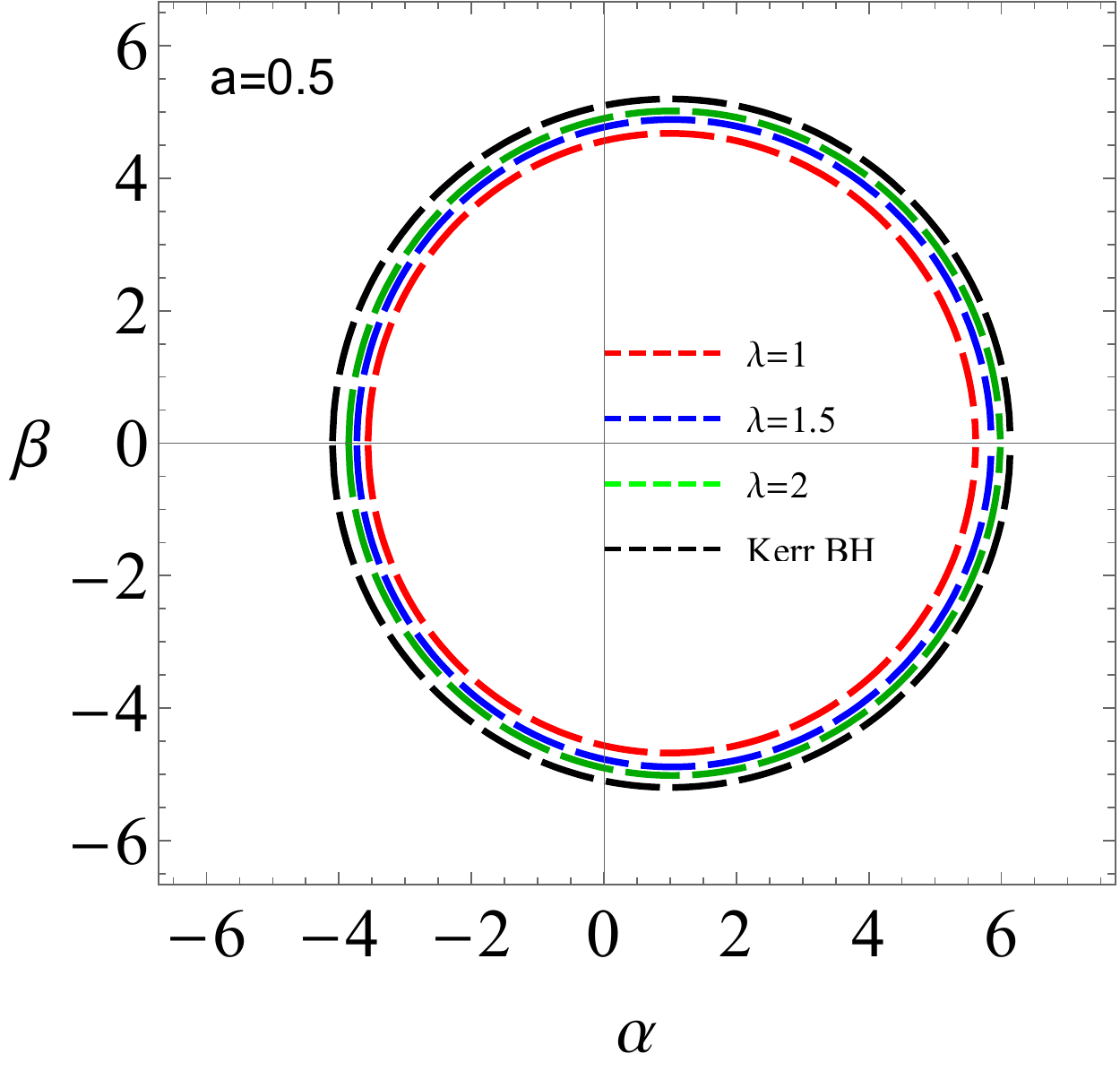}
\includegraphics[width=7.1cm]{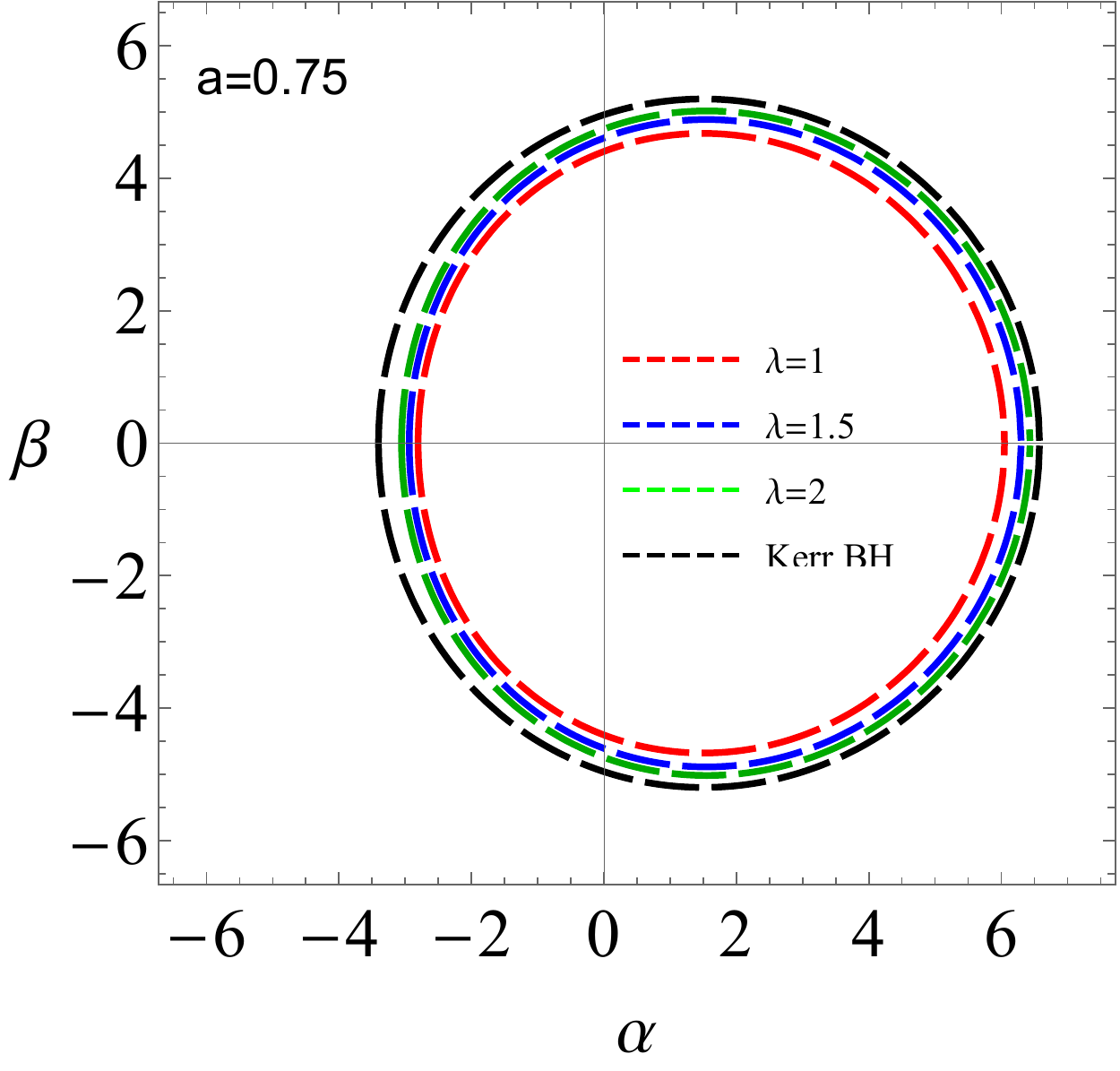}
\includegraphics[width=7.1cm]{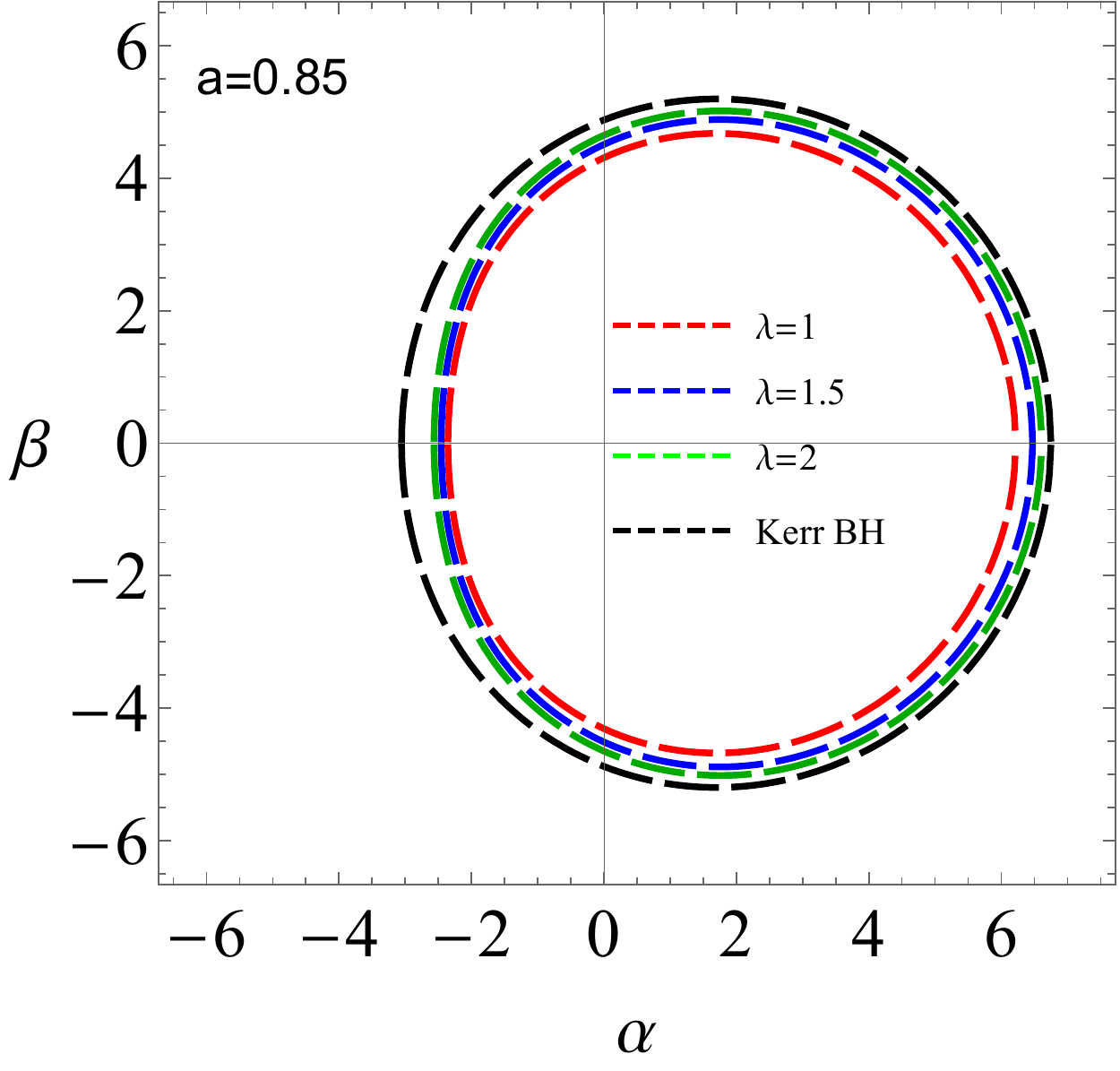}
\caption{The shape of shadow for $M=1$ and $Q=-0.2$ and postive $\lambda$.  } \label{B}
\end{figure*}

\begin{figure*}
\includegraphics[width=7.1cm]{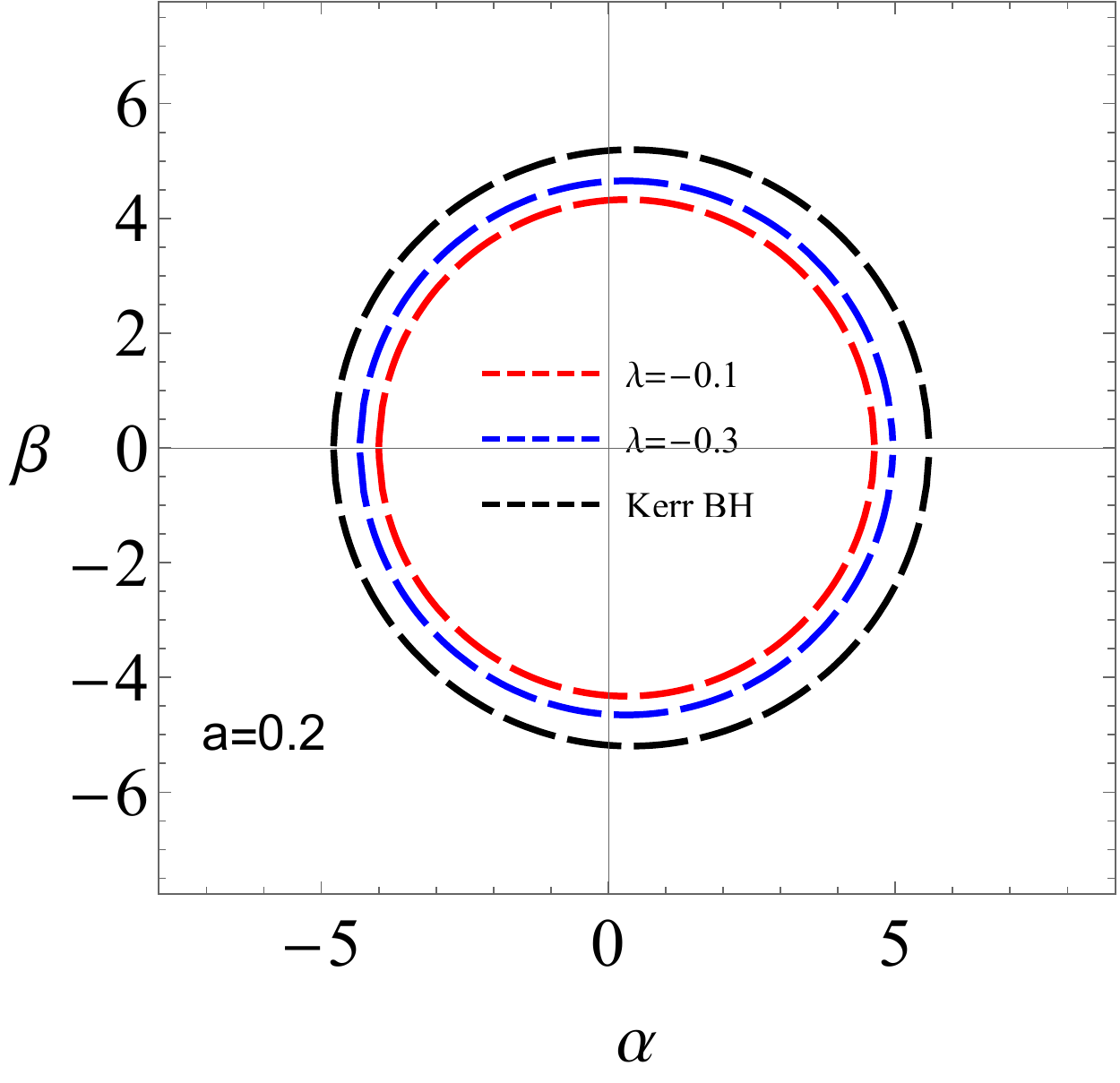}
\includegraphics[width=7.1cm]{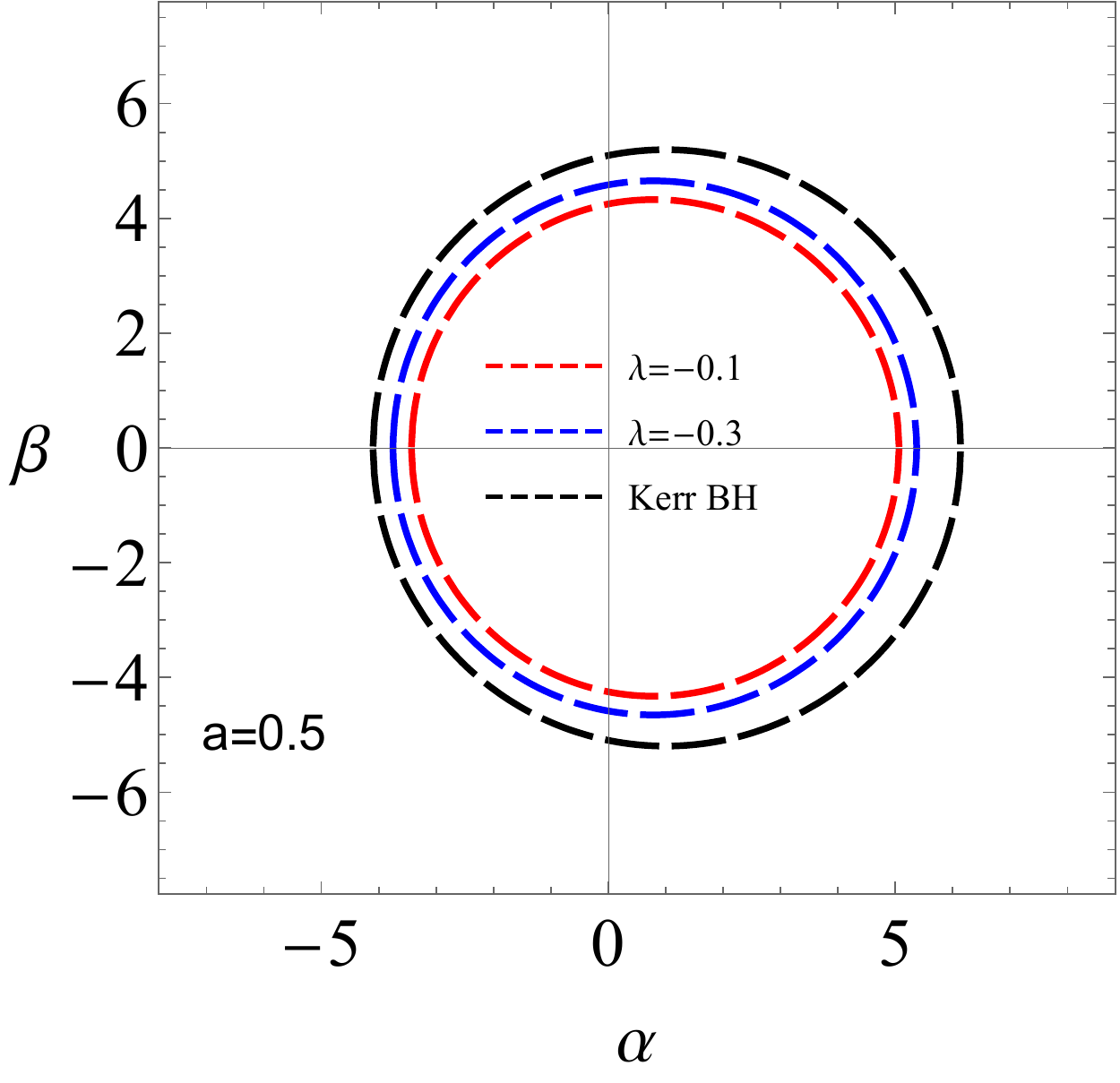}
\includegraphics[width=7.1cm]{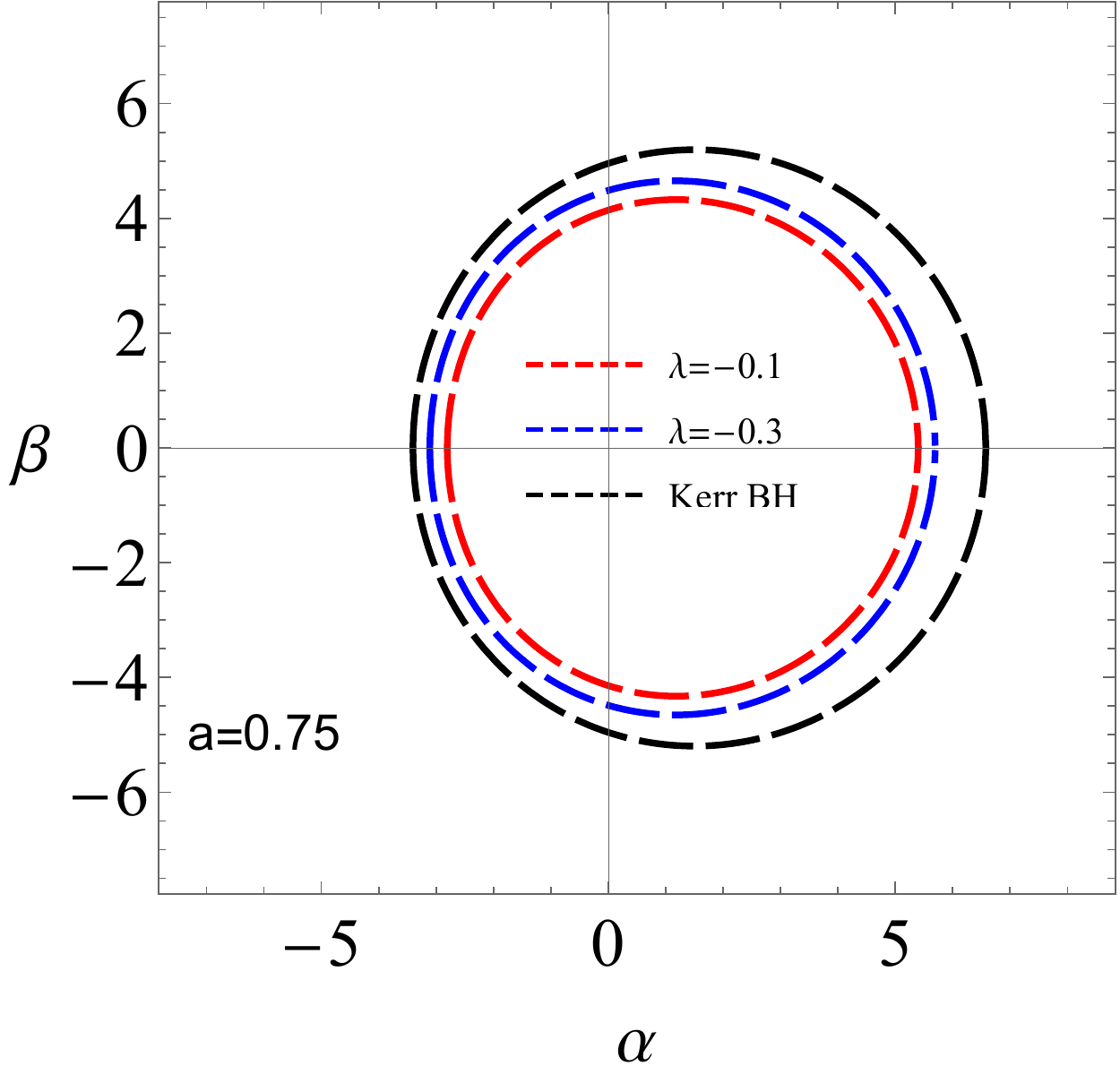}
\includegraphics[width=7.1cm]{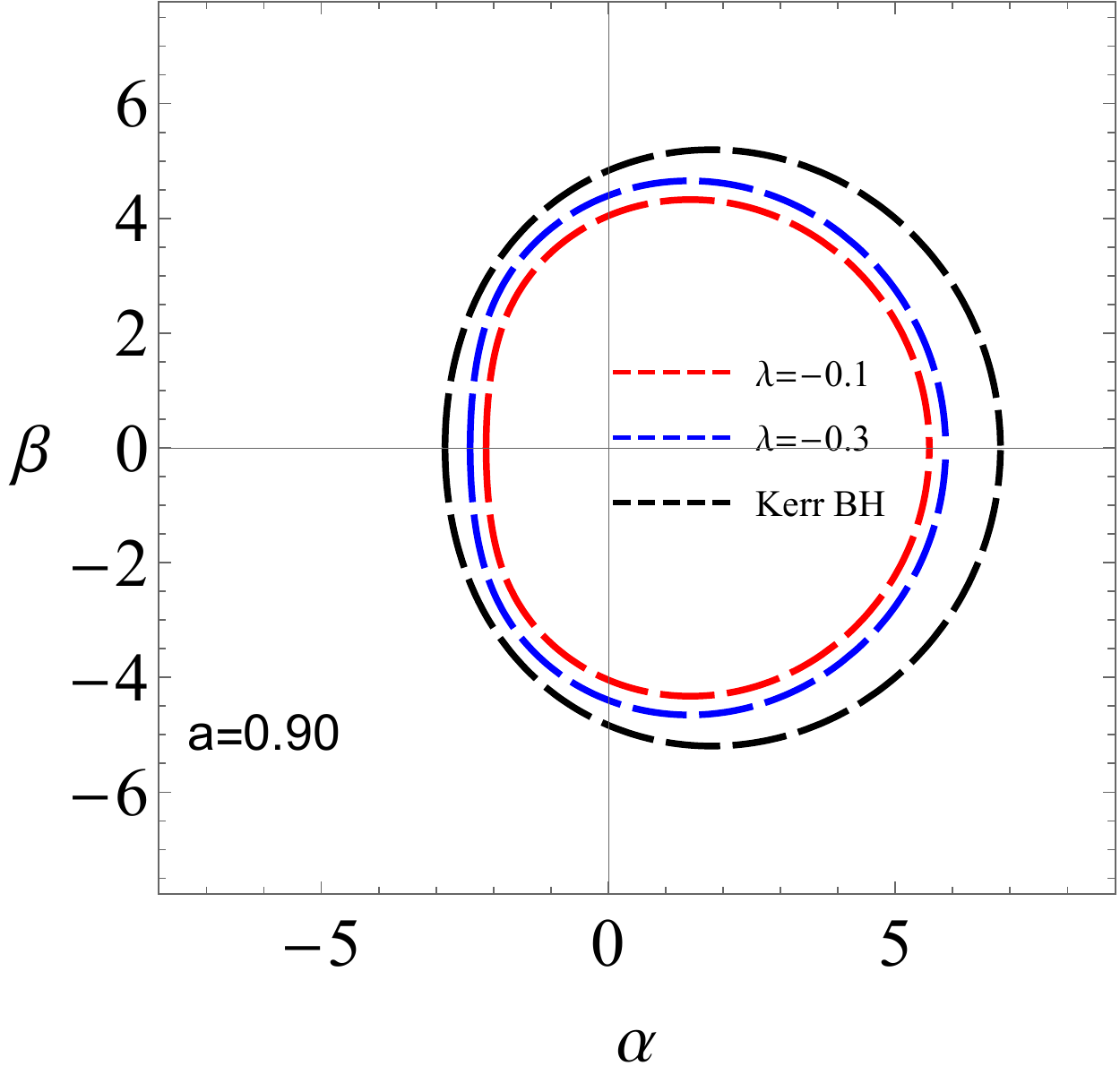}
\caption{The shape of shadow for $M=1$ and $Q=-0.2$ and negative $\lambda$.  } \label{B}
\end{figure*}

\begin{figure}
\includegraphics[width=7.1cm]{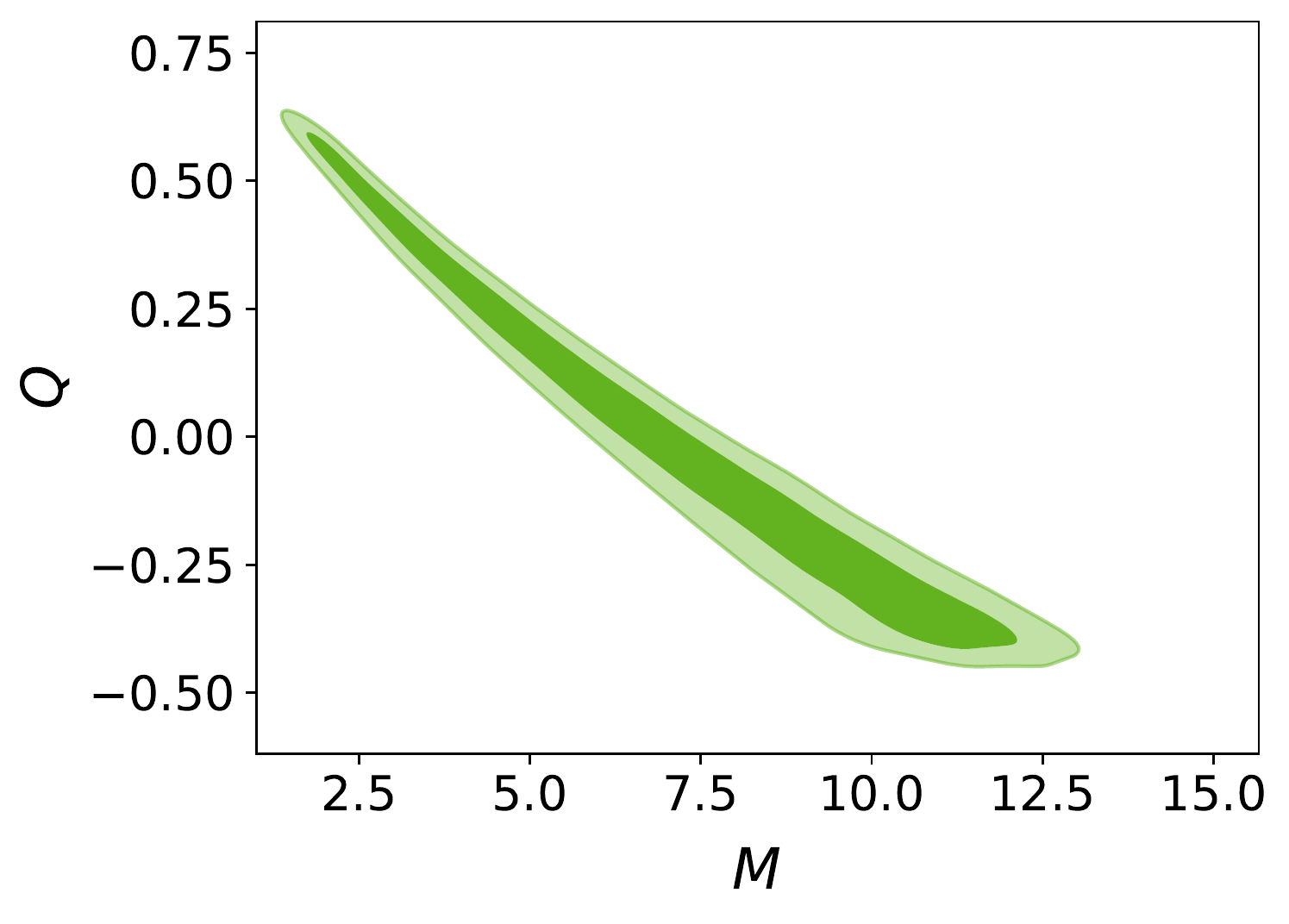}
\caption{Constraints on the parameter $Q$ and estimated M87* black hole mass $M(\times 10^9M_{\odot})$ using M87 shadow size in $68\%$ and $95\%$ confidence levels. \label{mq1}} 
\end{figure}

\section{Separation of null geodesic equations and black hole shadow}
\label{sec:shadow}

In this section, we consider the null geodesic equations in the general rotating spacetime (39) using the Hamilton-Jacobi method and obtain a general formula for finding the contour of a shadow. The Hamilton-Jacobi equation is given by
\begin{equation}
\frac{\partial \mathcal{S}}{\partial \sigma}=-\frac{1}{2}g^{\mu\nu}\frac{\partial \mathcal{S}}{\partial x^\mu}\frac{\partial \mathcal{S}}{\partial x^\nu},
\label{eq:HJE}
\end{equation}
where $\sigma$ is the affine parameter, $\mathcal{S}$ is the Jacobi action. There are two conserved quantities, the conserved energy $E=-p_t$ and the conserved angular momentum $J=p_\phi$ (about the axis of symmetry). In order to find a separable solution of Eq. (\ref{eq:HJE}), we can express the action in terms of the known constants of the motion as follows
\begin{equation}
\mathcal{S}=\frac{1}{2}\mu ^2 \sigma - E t + J \phi + \mathcal{S}_{r}(r)+\mathcal{S}_{\theta}(\theta),
\label{eq:action_ansatz}
\end{equation}
where $\mu$ is the mass of the test particle. For a photon, we take $\mu=0$. Putting Eq. (\ref{eq:action_ansatz}) in the Hamilton-Jacobi equation and using the Lagrangian
\begin{equation}
\mathcal{L}=\frac{1}{2}g_{\mu \nu}\dot{x}^{\mu}\dot{x}^{\nu},
\end{equation}
it is straightforward to recover the following equations of motions
\begin{equation}
W(r,\theta)\rho^2\frac{dt}{d\sigma}=\frac{r^2+a^2}{\Delta}[E(r^2+a^2)-aJ]-a(aE\sin^2\theta-J),
\end{equation}
\begin{equation}
W(r,\theta)\rho^2\frac{d\phi}{d\sigma}=\frac{a}{\Delta}[E(r^2+a^2)-aJ]-\left(aE-\frac{J}{\sin^2\theta}\right),
\end{equation}
\begin{equation}
W(r,\theta)H\frac{dr}{d\sigma}=\pm \sqrt{R(r)},
\label{eq:r_eqn}
\end{equation}
\begin{equation}
W(r,\theta)H\frac{d\theta}{d\sigma}=\pm \sqrt{\Theta(\theta)},
\label{eq:theta_eqn}
\end{equation}
where
\begin{equation}
R(r)=\left[X(r)E-aJ\right]^2-\Delta(r)\left[\mathcal{K}+\left(J-aE\right)^2\right],
\end{equation}
\begin{equation}
\Theta(\theta)=\mathcal{K}+a^2E^2\cos^2\theta-J^2\cot^2\theta,
\end{equation}
where $X(r)=(r^2+a^2)$, and $\Delta(r)$ is defined by Eq. (\ref{delta}), while $\mathcal{K}$ is known as the Carter separation constant. If we define $\xi=J/E$ and $\eta=\mathcal{K}/E^2$, for the unstable circular photon orbits one has the following conditions: $R(r_{ph})=0$, $R'(r_{ph})=0$ and $R''\geq 0$. Note that $r=r_{ph}$ gives the photon orbit radius. From these conditions it follows (see, \cite{Shaikh:2019fpu})
\begin{equation}
\left[X(r_{ph})-a\xi\right]^2-\Delta(r_{ph})\left[\eta+\left(\xi-a\right)^2\right]=0,
\label{eq:Req0}
\end{equation}
\begin{equation}
2X'(r_{ph})\left[X(r_{ph})-a\xi\right]-\Delta'(r_{ph})\left[\eta+\left(\xi-a\right)^2\right]=0.
\label{eq:Rpeq0}
\end{equation}

If we eliminate $\eta$ and solve for $\xi$ evaluated at $r=r_{ph}$, it follows \cite{Shaikh:2019fpu}
\begin{equation}
\xi=\frac{X_{ph}\Delta'_{ph}-2\Delta_{ph}X'_{ph}}{a\Delta'_{ph}},
\label{eq:xi}
\end{equation}
\begin{equation}
\eta=\frac{4a^2X'^2_{ph}\Delta_{ph}-\left[\left(X_{ph}-a^2\right)\Delta'_{ph}-2X'_{ph}\Delta_{ph} \right]^2}{a^2\Delta'^2_{ph}}.
\label{eq:eta}
\end{equation}
 Thus, equations (\ref{eq:xi}) and (\ref{eq:eta}) are basic equations to study the black hole shadow. To obtain the apparent shape of a black hole shadow, we need to introduce the celestial coordinates $\alpha$ and $\beta$ which by construction lie in the celestial plane perpendicular to the line joining the observer and the center of the spacetime geometry. These coordinates $\alpha$ and $\beta$ are defined as follows \cite{Hioki:2008zw}
\begin{equation}
\alpha=-r_o \frac{p^{(\phi)}}{p^{(t)}},\,\,\,\,\beta=r_o\frac{p^{(\theta)}}{p^{(t)}},
\end{equation}
in which $(p^{(t)},p^{(r)},p^{(\theta)},p^{(\phi)})$ are the tetrad components of
the photon momentum with respect to locally non-rotating reference frame. Furthermore one can write these coordinates in terms of $\xi$ and $\eta$, as follows  \cite{Kumar}
\begin{equation}
 \alpha = -r_o \dfrac{\xi}{\sqrt{g_{\phi \phi}}(\zeta-\gamma \xi)}\Big\vert_{(r_0,\theta_0)},\\\label{alphaa}
 \end{equation}
 \begin{equation}
\beta = \pm r_o \frac{\sqrt{\Theta_{\theta}(\theta)}}{\sqrt{g_{\theta \theta}(\zeta-\gamma \xi)}}\Big\vert_{(r_0,\theta_0)},\label{betaa}
\end{equation}
with
\begin{equation}
    \zeta=\sqrt{\frac{g_{\phi\phi}}{g_{t \phi}^2-g_{tt}g_{\phi \phi}}},\,\,\,\, \gamma=-\frac{g_{t \phi}}{g_{\phi \phi}}\zeta.
\end{equation}
Note that $(r_0,\theta_0)$ are the position coordinates of the observer.  For an observer sitting in the asymptotically flat region, we can take the limit $r_0 \to \infty$ to obtain
\begin{equation}
\alpha=-\frac{\xi}{\sin\theta_0},
\label{eq:alpha}
\end{equation}
\begin{equation}
\beta=\pm \sqrt{\eta+a^2\cos^2\theta_0-\xi^2\cot^2\theta_0},
\label{eq:beta}
\end{equation}
provided $\lambda>0$. The shadows are constructed by using the unstable photon orbit radius $r_{ph}$ as a parameter and then plotting parametric plots of $\alpha$ and $\beta$ using Eqs. (\ref{eq:xi}), (\ref{eq:eta}), (\ref{eq:alpha}) and (\ref{eq:beta}). However, when $\lambda<0$, the spacetime metric is not asymptotically flat, therefore we need to use Eqs. (\ref{alphaa}) and (\ref{betaa}).

\section{Shadow of regular black holes in massive gravity}

 Using the expressions for $\xi$ and $\eta$  we find the following results
\begin{equation}
\xi=\frac{2 r^\lambda (r^2(r-3M)+a^2(M+r))-r\Upsilon_1 Q}{2 a(M-r)r^\lambda-a (\lambda_1-2)r Q},
\end{equation}
\begin{equation}
\eta=\frac{8 a^2 r^{3+\lambda}(2 M r^\lambda+\lambda r Q)-r^4\Upsilon_2^2}{(-2 a (M-r)r^\lambda+a(\lambda-2)r Q)^2},
\end{equation}
where 
\begin{equation}
    \Upsilon_1=a^2(\lambda-2)+(\lambda+2)r^2,
\end{equation}
\begin{equation}
    \Upsilon_2=2(3M-r)r^\lambda+(2+\lambda)r Q.
\end{equation}

Depending on the sign before $Q$ and $\lambda$, we can have two interesting cases:
Firstly, when $\lambda>0$ or $\lambda<0$, provided $Q>0$, we observe that the size of the apparent shadow is bigger compared to that of the Kerr solution. Moreover with the decrease of $\lambda$ in absolute value while keeping $Q>0$, the shadow radius increases. Secondly, when $\lambda>0$ and $\lambda<0$, provided $Q<0$, we observe that the size of apparent shadows is smaller compared to that of the Kerr solution. That is, with a decrease of $\lambda$ in absolute value the shadow radius decreases.
One can observe a correspondence between the size of the event horizon area and the black hole shadow size. Namely, with the increase of the event horizon area (see Fig. 5), the shadow images also increase. On the other hand, if the event horizon area decreases (see Fig 6), the shadow radius also decreases. This is in agreement with a recent proposal \cite{Zhang:2019glo}, according to which there is a correspondence between the second law and black hole shadow. Namely, the event horizon area of the black hole will not
decrease with time 
\begin{equation*}
\delta A \geq 0,
\end{equation*}
and the shadow radius of the black hole does not decrease with time. This is related to the fact that the temperature of the Kerr black hole can be a function of not only the event horizon radius $r_h$, but also the
radius of the shadow $r_{s}$.

\subsection{Observational constraints }

We can apply our numerical results of shadow size to the latest observation of the black hole shadow of M87*. The first M87* Event Horizon Telescope (EHT) results published the image of shadow of black hole with a ring diameter of $42\pm3 \mu as$ \cite{EHT}. Adopting this measurement value and distance $D=16.8\pm0.8\text{Mpc}$, we did the Monte-Carlo simulations for the parameters space $(M,Q,\lambda)$. In $95\%$ condidence level, the charge parameter $Q$ is constrained as $Q=-0.05^{+0.52}_{-0.38} $.  The mass of M87* is estimated as $M=(7.6^{+4.5}_{-5.5} )\times 10^9M_{\odot}$ which is greater than the value derived by EHT $M=(6.5 \pm 0.7)\times 10^9 M_{\odot}$ in GR but with large uncertainty. The contour plot Fig(\ref{mq1}) shows the strong degeneracy between the parameter $Q$ and mass $M$ which is to be expected as they have leading contribution to shadow size. Meanwhile the parameter $\lambda$ can't be constrained well since it is not sensitive with size of shadow.

\section{Deflection angle of massive particles}

From the form of the metric (39), it is easy to see that the deflection angle of light should not depend upon the conformal factor. Therefore, let us now focus on a more interesting problem, namely the gravitational deflection of relativistic massive particles using the Gauss-Bonnet theorem and following the approach introduced by Crisnejo, Gallo and Jusufi \cite{Crisnejo:2019ril}.  For a given stationary, and axisymmetric spacetime $(\mathcal{M},g_{\alpha\beta})$, in the presence of cold non-magnetized plasma with the refractive index $n$ is given by
\begin{equation}\label{refra-index}
    n^2(x,\omega(x))=1-\frac{\omega_e^2(x)}{\omega^2(x)},
\end{equation}
where $\omega(x)$ gives the photon frequency measured by an observer following a timelike Killing vector field. On the other hand, $\omega_e(x)$, is known as the plasma frequency given by
\begin{equation}\label{K_e}
    \omega_e^2(x)=\frac{4\pi e^2}{m_e} N(x)= K_e N(x),
\end{equation}
here $e$ and $m_e$ represents the charge and the mass of the electron, respectively. Furthermore $N(x)$ gives the number density of electrons in the plasma media. It is worth noting that the quantity $\omega(x)$ in terms of the gravitational redshift is given by
\begin{equation}
    \omega(x)=\frac{\omega_\infty}{\sqrt{-g_{00}}}.
\end{equation}

To calculate the deflection angle of massive particles, we shall use a correspondence between the motion of a photon in a cold non-magnetized plasma and the motion of a test massive particle in the same background.
In particular, one should identify the electron frequency of the plasma  $\hbar\omega_e$ with the mass $m$ of the test massive particle and the total energy $E=\hbar \omega_\infty$ of a photon with the total energy $E_\infty=m/(1-v^2)^{1/2}$. With those pieces of information in hand, let us formulate the Gauss-Bonnet theorem. 

\textbf{Theorem}. \textit{Let $D\subset S$ be a regular domain of an oriented two-dimensional surface $S$ with a Riemannian metric $\tilde{g}_{ij}$, the boundary of which is formed by a closed, simple, piece-wise, regular, and positive oriented curve $\partial D: \mathcal{R}\supset I\to D$. Then,}
\begin{equation}
    \int\int_D \mathcal{K}dS+\int_{\partial D} k_g dl +\sum_i \epsilon_i = 2\pi\chi(D), \ \ \ \sigma\in I.
\end{equation}

Note that $\chi(D)$ and $\mathcal{K}$ are known as the Euler characteristic and Gaussian
curvature of the optical domain $D$, respectively. Moreover $k_g$ represents the geodesic curvature of the optical domain $\partial D$, while, $\epsilon_i$ gives the corresponding exterior angle in the i-th vertex.  Without going into details,  the Gauss-Bonnet theorem can be reformulated in terms of the deflection angle $\hat{\alpha}$, as follows \cite{Crisnejo:2019ril}
\begin{equation}
    \int_0^{\pi+\hat{\alpha}}\bigg[\kappa_g \frac{d\sigma}{d\phi}\bigg]\bigg|_{C_R}d\phi=\pi -\bigg(\int\int_{D_R}\mathcal{K}dS + \int_{\gamma_p} k_g dl \bigg),
\end{equation}
where the limit $R\to\infty$ was applied.
For the asymptotically flat spacetimes,  one can easily show the following condition
$[k_g\frac{d\sigma}{d\phi}]_{C_R}\to 1$ when the radius of $C_R$ tends to infinity. This in turn from the GBT allows us to express the deflection angle by the simple relation 
\begin{equation}\label{alpha-general}
    \hat{\alpha}=-\int\int_{D_R}\mathcal{K}dS - \int_{\gamma_p} k_g dl.
\end{equation}
In this work we will restrict our attention to asymptotically flat spacetimes, and by then the expression \eqref{alpha-general} will be enough to calculate the deflection angle.
In particular, for light rays moving in the equatorial plane characterized by $\theta=\pi/2$, its geodesic curvature can be calculated as follows,
\begin{equation}\label{eq:kgasada}
    k_g =-\frac{1}{\sqrt{\hat{g} \hat{g}^{\theta\theta} }} \partial_r\hat{\beta}_\phi,
\end{equation}
where $\hat{g}$ is the determinant of $\hat{g}_{ab}$. Next, the metric on the equatorial plane can be written as,
\begin{equation}\label{eq:metricoriginal}
\begin{split}
    ds^2&=-A\,dt^2+B\,dr^2-2Hdtd\phi +D d\phi^2.
    \end{split}
\end{equation}
By making use of the above correspondence on the metric, we find the Finsler-Randers metric determined by
\begin{equation}
    \mathcal{F}(x,\dot{x})=\sqrt{\hat{g}_{ab} \dot{x}^a \dot{x}^b}+ \bm{\hat\beta}_a \dot{x}^{a},
\end{equation}
where
\begin{eqnarray}
    \hat{g}_{ab} dx^a dx^b&=&n^2[(\frac{B}{A})dr^2\nonumber+\frac{AD+H^2}{A^2}d\phi^2],\nonumber\\
    \bm{\hat\beta}&=&-\frac{H}{A}d\phi.
\end{eqnarray}
Furthermore, in the linear order of $a$ one has
\begin{eqnarray}
    A(r)&=& W(r)(1-\frac{2Mr+Q r^{2-\lambda}}{r^2}),\\
    B(r)&=& \frac{W(r)}{(1-\frac{2Mr+Q r^{2-\lambda}}{r^2})},\\
    D(r)&=& W(r) r^2,\\
    H(r)&=& W(r) \frac{2 a M}{r},
\end{eqnarray}
along with the refractive index given by
\begin{eqnarray}
    n^2(r)&=& 1-(1-v^2)A(r).
\end{eqnarray}
With this identification at hand, the deflection angle is expressed as follows
\begin{equation}\label{alpha-pm}
    \hat{\alpha}_{\text{mp}}=-\iint_{D_r}\mathcal{K}dS-\int_{R}^S k_g dl.
\end{equation}
Note that $l$ is some affine parameter (see, \cite{Crisnejo:2019ril}). In addition, $S$ represents the source and $R$ represents the receiver, respectively. In what follows we will consider some special cases depending on the value of $\lambda$, provided $\lambda>0$. Otherwise, for a general $\lambda$, one can only study finite distance corrections. 

\subsubsection{Case $\lambda=1$}
As a first example, we will consider the case $\lambda=1$. Moreover, our metric (39) at the first order in $a$, and by restricting our attention to the equatorial plane yields
\begin{equation}
\begin{aligned}
    ds^2 =\, &W(r)\Big[-(1-\frac{2M}{r}-\frac{Q}{r})dt^2-2a \left(\frac{2M}{r}\right) dt d\phi \\
    &+ \frac{dr^2}{(1-\frac{2M}{r}-\frac{Q}{r})} + r^2d\phi^2\Big].
\end{aligned}    
\end{equation}
The Gaussian optical curvature is found to be
\begin{equation}
    \mathcal{K}\simeq -\frac{M}{r^3 v^2 }(1+\frac{1}{v^2})-\frac{Q}{2 r^3 v^2 }(1+\frac{1}{v^2})+\frac{10 L^2}{r^4 v^2 }(1-\frac{1}{v^2}).
\end{equation}
On the other hand for the geodesic curvature contribution we find
\begin{equation}
    \Big[k_gdl\Big]_{r_{\gamma}} \simeq -\frac{2 a M}{v b^2}\sin\phi d\phi.
\end{equation}
The deflection angle reads
\begin{equation}\label{alpha-pm}
   \hat{\alpha}_{\text{mp}}=-\int_{0}^{\pi}\int_{b/\sin(\phi)}^{\infty}\mathcal{K} \sqrt{\det \hat{g}} dr d\phi-\int_{0}^{\pi}s\frac{2 a M}{v b^2}\sin\phi d\phi,
\end{equation}
with $s=+1$ for prograde orbits and $s=-1$ for retrograde ones. Utilizaing our expression \eqref{alpha-pm} we obtain the following result for the deflection angle
\begin{eqnarray}\label{eq:eRN}\notag
    \hat{\alpha}_{\text{mp}} &=& \frac{2M}{b}\bigg(1+\frac{1}{v^2}\bigg)+\frac{Q}{b}\bigg(1+\frac{1}{v^2}\bigg)-\frac{5 \pi L^2}{4 b^2}(1-\frac{1}{v^2})\\
    &-& \frac{4saM}{b^2 v},
\end{eqnarray}
where $s$ stands for the prograde/retrograde orbit. Furthermore setting $v=1$, we obtain the deflection angle of light given by (see, \cite{Jusufi:2017drg})
\begin{eqnarray}
    \hat{\alpha}_{\text{light}} &=& \frac{4M}{b}+\frac{2Q}{b}- \frac{4saM}{b^2}.
\end{eqnarray}

Thus, as we expected, there is no effect of $L^2$ on the light deflection. The corresponding result for the static case was previously reported in Ref. \cite{Jusufi:2017drg}.

\subsubsection{Case $\lambda=2$}
In this case, the linearized metric can be written as follows
\begin{equation}
\begin{aligned}
    ds^2 =\, &W(r)\Big[-(1-\frac{2M}{r}-\frac{Q}{r^2})dt^2-2a \left(\frac{2M}{r}\right) dt d\phi \\
    &+ \frac{dr^2}{(1-\frac{2M}{r}-\frac{Q}{r^2})} + r^2d\phi^2\Big].
\end{aligned}    
\end{equation}
The  Gaussian optical curvature is found to be
\begin{equation}
    \mathcal{K}\simeq -\frac{M}{r^3 v^2 }(1+\frac{1}{v^2})-\frac{2 Q}{r^4 v^2 }(1+\frac{2}{v^2})+\frac{12 L^2}{r^4 v^2 }(1-\frac{1}{v^2}).
\end{equation}
Using  \eqref{alpha-pm} we obtain the following result for the deflection angle
\begin{eqnarray}\notag
    \hat{\alpha}_{\text{mp}} &=& \frac{2M}{b}\bigg(1+\frac{1}{v^2}\bigg)+\frac{\pi Q}{4b^2}\bigg(1+\frac{2}{v^2}\bigg)-\frac{3 \pi L^2}{2 b^2}(1-\frac{1}{v^2})\\
    &-& \frac{4saM}{b^2 v}.
\end{eqnarray}
As a special case, $Q\to -Q^2$ and $L^2=0$, we find the deflection angle in a Kerr-Newman spacetime obtained in  \cite{Crisnejo:2019ril}. The deflection angle of light follows the limit $v=1$, yielding \cite{Jusufi:2017drg}
\begin{eqnarray}
    \hat{\alpha}_{\text{light}} &=& \frac{4M}{b}+\frac{3\pi Q}{4b^2}- \frac{4saM}{b^2}.
\end{eqnarray}

\subsubsection{Case $\lambda=3$}
As the last example we shall consider $\lambda=3$. The spacetime metric in this case reads
\begin{equation}
\begin{aligned}
    ds^2 =\, &W(r)\Big[-(1-\frac{2M}{r}-\frac{Q}{r^3})dt^2-2a \left(\frac{2M}{r}\right) dt d\phi \\
    &+ \frac{dr^2}{(1-\frac{2M}{r}-\frac{Q}{r^3})} + r^2d\phi^2\Big].
\end{aligned}    
\end{equation}
The Gaussian optical curvature  reads 
\begin{equation}
    \mathcal{K}\simeq -\frac{M}{r^3 v^2 }(1+\frac{1}{v^2})-\frac{3Q}{r^5 v^2 }(1+\frac{3}{v^2})+\frac{14 L^2}{r^4 v^2 }(1-\frac{1}{v^2}).
\end{equation}
Using the expression \eqref{alpha-pm} we obtain the deflection angle for prograde/retrograde orbits of massive particles,
\begin{eqnarray}\notag
    \hat{\alpha}_{\text{mp}} &=& \frac{2M}{b}\bigg(1+\frac{1}{v^2}\bigg)+\frac{2Q}{3b^3}\bigg(1+\frac{3}{v^2}\bigg)-\frac{7 \pi L^2}{4 b^2}(1-\frac{1}{v^2})\\
    &-& \frac{4saM}{b^2 v}.
\end{eqnarray}
Finally, the deflection angle of light results with
\begin{eqnarray}
    \hat{\alpha}_{\text{light}} &=& \frac{4M}{b}+\frac{8Q}{3b^3}- \frac{4saM}{b^2 v}.
\end{eqnarray}
For the static spacetime the corresponding result was obtained by Jusufi et al. \cite{Jusufi:2017drg}.

\section{Conclusion}
In this paper, we have constructed static and rotating regular black holes in conformal massive gravity. Choosing a suitable scaling factor, we have shown that the scalar invariants are indeed regular everywhere. Moreover, to justify the regularity of the spacetime, we have checked the geodesics completeness for a test massive particle by calculating the proper time. To this end, we have investigated in more detail the shadow images and the deflection angle of relativistic massive particles in the spacetime geometry of the regular rotating black hole. Our result shows that:
\begin{itemize}
    \item For $\lambda>0$, or $\lambda<0$, and $Q>0$,
\end{itemize}
 the shadow images are larger compared to the Kerr vacuum black hole shadow. Specifically, with the decrease of $\lambda$ in absolute value the shadow radius increases. 
 \begin{itemize}
    \item For $\lambda>0$, or $\lambda<0$, and $Q<0$.
\end{itemize}
 the size of the shadows is smaller compared to the Kerr vacuum black hole shadow. In this case, with the decrease of $\lambda$ in absolute value the shadow radius decreases, compared to the Kerr black hole shadow. We also put observational constraints on the parameter $ Q $ using the latest EHT observation of the supermassive black hole M87*. Finally, using the correspondence between the motion of a photon in a cold non-magnetized plasma and the motion of a test massive particle in the same background and utilizing the Gauss-Bonnet theorem over the optical geometry, we have calculated the deflection angle of massive particles. Among other things, we have shown that the deflection angle is affected in leading order by the conformal factor. In fact, the choice of conformal factor plays a crucial role in calculating the deflection angle. This important result shows that the deflection angle of particles can be used to distinguish a rotating regular black hole from a rotating singular black hole. As a special case when $v=1$, the deflection angle of light is obtained. As expected, the conformal factor does not affect the deflection angle.

\section*{Acknowledgement}
H.C. would like to acknowledge the hospitality of ITPC, Zhejiang University of Technology, Hangzhou where part of this work was started. H.C. also acknowledges support from the China Scholarship Council (CSC), grant No.~2017GXZ019020. The work of A.W. was supported in part by   the National Natural Science Foundation of China (NNSFC), Grant Nos. 11675145, and 11975203.

\appendix
\section{}\label{apend_a}
The Ricci and Kretchmann scalars read, respectively, as 
\begin{widetext}
	\begin{equation}
	\begin{aligned}
	& \hat{\mathcal{R}} = -\frac{r^{1-\lambda } \left(\frac{L^2}{r^2}+1\right)^{-\frac{\left| \lambda \right|}{2}}}{2 \left(L^2+r^2\right)^4} \Bigg[3 L^4 \left| \lambda \right| ^2 \left((r-2 M) r^{\lambda }-Q r\right) \\
	& \ \ \ \ +6 L^2 \left| \lambda \right|  \Big(L^2 \left((3 r-8 M) r^{\lambda }-(\lambda +3) Q r\right) \\
	& \ \ \ \ +r^2 \left((r-4 M) r^{\lambda }-(\lambda +1) Q r\right)\Big) \\
	& \ \ \ \ +2 \left(L^2+r^2\right) \Big(L^2 \big(-12 (4 M-r) r^{\lambda } \\ 
	& \ \ \ \ -\left(\lambda ^2+9 \lambda +14\right) Q
	r\big)-\left(\lambda ^2-3 \lambda +2\right) Q r^3\Big)\Bigg],
	\end{aligned}
	\end{equation}
and
	\begin{equation}
		\begin{aligned}
			&\hat{K} = \frac{r^{2-2 \lambda } \left(\frac{L^2}{r^2}+1\right)^{-\left| \lambda \right| }}{4 \left(L^2+r^2\right)^8}\Bigg[ 3 \left((2 M-r) r^{\lambda }+Q r\right)^2 \left| \lambda \right| ^4 L^8+ \\
			& 4 \left(r^{\lambda } (r-2 M)-Q r\right) \left(2 (3 M-2 r) r^{\lambda +2}-Q (\lambda -4) r^3+L^2 \left(-2 (9 M-4 r) r^{\lambda }-Q (\lambda +8) r\right)\right) \left| \lambda \right| ^3 L^6 \\
			& +4 \Big(2 Q (2 M (5 \lambda +28)-3 r (\lambda +8)) r^{\lambda +5}+\left(132 M^2-108 r M+23 r^2\right) r^{2 \lambda +4}+Q^2 \left(2
			\lambda ^2+6 \lambda +25\right) r^6 \\
			& +L^2 \left(8 Q (6 \lambda  M-9 M+7 r-2 r \lambda ) r^{\lambda +3}-2 \left(12 M^2-40 r M+15 r^2\right) r^{2 \lambda +2}+2 Q^2 \left(2 \lambda ^2+8 \lambda -13\right) r^4\right) \\
			& +L^4 \left(\left(180 M^2-148 r M+31 r^2\right) r^{2 \lambda }+2 Q (14 \lambda  M+76 M-32 r-5 r \lambda ) r^{\lambda +1}+Q^2 \left(2 \lambda ^2+10 \lambda +33\right) r^2\right)\Big) \left| \lambda \right| ^2 L^4 \\
			& +4 \Big(8 M Q \left(\lambda ^2+3 \lambda +2\right) r^{\lambda +9}+48 M^2 r^{2 \lambda +8}+Q^2 \left(\lambda ^4+2 \lambda ^3+5 \lambda ^2+4\right) r^{10} \\
			& +4 L^2 \left(48 M^2 r^{2 \lambda }-2 Q \left(r \left(3 \lambda ^2+\lambda -4\right)-12 M \lambda  (\lambda +1)\right) r^{\lambda +1}+Q^2 \left(\lambda ^4+4 \lambda ^3+9 \lambda ^2+2 \lambda -4\right) r^2\right) r^6 \\
			& +2 L^4 \Big(8 \left(150 M^2-108 r M+23 r^2\right) r^{2 \lambda }+4 Q \left(M \left(26 \lambda ^2+62 \lambda +212\right)-r \left(7 \lambda ^2+17 \lambda +84\right)\right) r^{\lambda +1} \\
			& +Q^2 \left(3 \lambda ^4+18 \lambda ^3+47 \lambda ^2+68 \lambda +164\right) r^2\Big) r^4+4 L^6 \Big(8 \left(42 M^2-16 r M+r^2\right) r^{2 \lambda } \\
			& +2 Q \left(4 M \left(5 \lambda ^2+21 \lambda +16\right)-r \left(5 \lambda ^2+23 \lambda +4\right)\right) r^{\lambda +1}+Q^2 \left(\lambda ^4+8 \lambda ^3+25 \lambda ^2+46 \lambda +4\right) r^2\Big) r^2 \\
			& +L^8 \Big(16 \left(39 M^2-20 r M+3 r^2\right) r^{2 \lambda }-8 Q \left(r \left(\lambda ^2+7 \lambda +12\right)-M \left(5 \lambda ^2+31
			\lambda +42\right)\right) r^{\lambda +1} \\
			& +Q^2 \left(\lambda ^4+10 \lambda ^3+37 \lambda ^2+56 \lambda +52\right) r^2\Big)\Big)+8 \Big(\Big(4 \left(60 M^2-40 r M+7 r^2\right) r^{2 \lambda }+Q \big(4 M \left(\lambda ^2+17 \lambda +42\right) \\
			& -r \left(\lambda ^2+19 \lambda +60\right)\Big) r^{\lambda +1}+Q^2
			\left(\lambda ^3+8 \lambda ^2+19 \lambda +32\right) r^2\Big) L^8+\Big(Q \Big(r \left(-5 \lambda ^2-47 \lambda +44\right) \\
			& +8 M \left(2 \lambda ^2+19 \lambda +3\right)\Big) r^{\lambda +3}-8 \left(-24 M^2+2 r M+3 r^2\right) r^{2 \lambda +2}+Q^2 \left(3 \lambda ^3+18 \lambda ^2+47 \lambda -20\right) r^4\Big) L^6 \\
			& +\Big(Q \left(4 M \left(5 \lambda ^2+21 \lambda +106\right)-r \left(7 \lambda ^2+29 \lambda +180\right)\right) r^{\lambda +5}+4 \left(132 M^2-108 r M+23 r^2\right) r^{2 \lambda +4} \\
			& +Q^2 \left(3 \lambda ^3+12 \lambda ^2+29 \lambda
			+88\right) r^6\Big) L^4+Q r^7 (\lambda -1) \left((8 M (\lambda +1)-r (3 \lambda +4)) r^{\lambda }+Q \left(\lambda ^2+3 \lambda +4\right) r\right) L^2\Big) \left| \lambda \right| \Bigg].
		\end{aligned}
	\end{equation}
\end{widetext}

\end{document}